\documentclass[12pt]{article}
\renewcommand{\baselinestretch}{2.2}
\usepackage{amsmath,epsfig,epsf,psfrag,natbib}
\usepackage{amssymb}
\usepackage{amsbsy}
\usepackage{lscape}
\usepackage{url,mathabx,bbm}
\usepackage{epstopdf}
\usepackage{color,latexsym,amsfonts, amsthm,amscd,graphicx}
\usepackage[varg]{txfonts}
\usepackage[margin=2cm]{geometry}

\usepackage[latin1]{inputenc}
\newcommand{\blind}{1}

\usepackage{amsbsy,lscape,epstopdf}
\usepackage{color,latexsym,amsfonts,bbm,amscd}

\newtheorem{theorem}{Theorem}

\newcommand{\bas}{\begin{eqnarray*}}
\newcommand{\eas}{\end{eqnarray*}}
\newcommand{\ba}{\begin{eqnarray}}
\newcommand{\ea}{\end{eqnarray}}
\newcommand{\bit}{\begin{itemize}}
\newcommand{\eit}{\end{itemize}}
\newcommand{\ben}{\begin{enumerate}}
\newcommand{\een}{\end{enumerate}}
\newcommand{\e}{ { \mathbb{E}}}

\newcommand{\bq}{\mbox{\bf q}}
\newcommand{\diag}{\mbox{diag}}

\newcommand{\bX}{\mbox{\bf X}}
\newcommand{\bV}{\mbox{\bf V}}
\newcommand{\bY}{\mbox{\bf Y}}

\newcommand{\bx}{\mbox{\bf x}}

\newcommand{\argmax}{{\mbox{argmax}}}
\newcommand{\amse}{{\mbox{\sc amse}}}

\newcommand{\bbeta}{\mbox{$\boldsymbol{\betaup}$}}
\newcommand{\ind}{\mathbbm{1}}

\newcommand{\btheta}{\mbox{$\boldsymbol{\thetaup}$}}
\newcommand{\bOmega}{\mbox{$\mathbf{\Omega}$}}
\newcommand{\bSigma}{\mbox{$\mathbf{\Sigma}$}}

\newcommand{\bh}{\mbox{\bf h}}

\newcommand{\bH}{\mbox{\bf H}}

\newcommand{\0}{\mbox{\bf 0}}

\newcommand{\converge}{\overset{d}{\longrightarrow}}

\newcommand{\ttin}{ \mbox{ttin}}
\newcommand{\gender}{ \mbox{gender}}
\newcommand{\age}{ \mbox{age}}

\newcommand{\yrx}{ \mbox{yrx}}
\newcommand{\tweek}{ \mbox{tweek}}
\newcommand{\edu}{ \mbox{edu}}

\begin{document}

\date{}
\title{\bf Small Area Quantile Estimation}

\def\spacingset#1{\renewcommand{\baselinestretch}%
{#1}\small\normalsize} \spacingset{1.3}

\if1\blind
{
  \title{\bf Small Area Quantile Estimation}
  \author{Jiahua Chen\thanks{
    The authors gratefully acknowledge funding from the ``a thousand talents'' program
    of Yunnan University and from NSERC Grant RGPIN-2014-03743,
    the National Natural Science Foundation of China (Number  11371142), the Program of
    Shanghai Subject Chief Scientist (14XD1401600), and  the 111 Project (B14019).}\\
    Research Institute of Big Data, University of Yunnan, China\\
    Department of Statistics, University of
    British Columbia, Canada \\
    Yukun Liu \\
    School of Statistics, East China Normal University, China  }
  \maketitle
} \fi

\begin{abstract}
Sample surveys are widely used to obtain information about
totals, means, medians, and other parameters of finite populations.
In many applications, similar information is desired
for subpopulations such as individuals in
specific geographic areas and socio-demographic groups.
When the surveys are conducted at national or similarly high levels,
a probability sampling can result in just a few sampling units from many
unplanned subpopulations at the design stage.
Cost considerations may also lead to low sample sizes from
individual small areas.
Estimating the parameters of these
subpopulations with satisfactory precision
and evaluating their accuracy
are serious challenges for statisticians.
To overcome the difficulties,
statisticians resort to pooling information across
the small areas via suitable model assumptions,
administrative archives, and census data.
In this paper, we develop an array of small area quantile estimators.
The novelty is the introduction
of a semiparametric density ratio model for the
error distribution in the unit-level nested error regression model.
In contrast, the existing methods are usually most effective when the
response values are jointly normal.
We also propose a resampling procedure for estimating
the mean square errors of these estimators.
Simulation results indicate that the new methods
have superior performance when the population distributions
are skewed and remain competitive otherwise.
\end{abstract}

\section{Introduction \label{intro}}

Sample surveys are widely used to obtain information about
the totals, means, medians, and other parameters of finite populations.
In many applications, the same information is desired
for subpopulations such as individuals in
specific geographic areas or in socio-demographic groups.
The estimation of finite subpopulation parameters is
referred to as the small area estimation problem (Rao  2003).
While the geographic areas may not be small, there may be a shortage of
direct information from individual areas.
Often, the surveys are conducted at national or similarly high levels.
The random nature of probability sampling can result in
just a few sampling units from many unplanned subpopulations
that are not considered at the design stage.
Cost considerations can also lead to low sample sizes.
Estimating the parameters of these
subpopulations with satisfactory precision
and evaluating their accuracy
are serious challenges for statisticians.

Because of the scarcity of direct information from small areas,
reliable estimates are possible only if indirect information
from other areas is available and effectively utilized.
This leads to a common thread of ``borrowing strength.''
Statisticians also seek auxiliary information
from sources such as administrative archives and census data
on subpopulations to obtain indirect estimates
for the subpopulation parameter.
These estimates may then be combined ``optimally.''

The small area estimation problem has been
intensively studied for many years.
Early publications covering foundational work
include Fay and Herriot (1979), Battese, Harter, and Fuller (1988),
Prasad and Rao (1990), and Lahiri and Rao (1995).
Successful applications can be found in
Schaible (1993), Tzavidis et al.~(2008),  and
Kriegler and Berk (2010).
Elbers, Lanjouw, and Lanjouw (2003) use a unit-level model that
combines census and survey data. The method has been employed
by many to reveal the spatial distribution of poverty and income inequality
(Haslett and Jones 2005; Neri, Ballini, and Betti 2005;
Ballini, Betti, Carrette, and Neri 2006; Tarozzi and Deaton 2009).
There are many papers containing
novel developments in theory and methodology; see
You  and Rao (2002),
Jiang and Lahiri (2006),
Pfeffermann and Sverchkov (2007),
Ghosh, Maiti, and Roy (2008),
Jiang, Nguyen, and Rao (2010),
Chaudhuri and Ghosh (2011),
Marchetti, Tzavidis, and Pratesi (2012),
Jiongo, Haziza, and Duchesne (2013), and
Verret, Rao, and Hiridoglou (2015).
We recommend Pfeffermann (2002, 2013), Rao (2003),
and Rao and Molina (2015) as additional references.

In this paper, we develop an array of new small area quantile estimators.
The existing methods such as that
proposed by Molina and Rao (2010) utilize optimal
prediction via the conditional expectation. This computation is
most convenient when the response values are jointly normal.
There are many ways to extend the approach to non-normal
data, e.g., transforming the response to improve the
fitness of the normal model or employing a skewed normal distribution
to compute the optimal predictions.
The novelty in our development is the introduction
of a semiparametric density ratio model for the
error distribution in the unit-level nested error regression model.
We avoid restrictive parametric assumptions
while ``borrowing strength'' between small areas.
We also propose a resampling procedure to estimate
the mean square errors of these estimators.
Our simulation results indicate that the new methods
have superior performance when the population distributions
are skewed and remain competitive otherwise.

The paper is organized as follows.
In Section 2, we review closely related developments.
In Section 3, we introduce the new methods.
In Section 4, we develop a resampling method for the estimation
of the mean square errors.
In Section 5, we give some theoretical results, leaving the technical proofs to the Appendix.
In Section 6, we use simulation to
reveal the properties of the new methods and compare them with
existing methods using artificial data sets and a real data set.
We end the paper with a summary and discussion.

\section{Literature review}
Let $\{(\bx_{kj}, y_{kj}): k=0, \ldots, m; j=1, \ldots, n_k\}$
be a random sample from a finite population with $m+1$ small
areas where the $k$th area contains $N_k$ sampling units.
We use $s_k$ to
denote the set of  observed sampling units in small area $k$.
We refer to $\bx_{kj}$ as an auxiliary variable.
In some applications, all the $\bx_{kj}$ values in the population
are available from a census or register.
In other applications, these values are known only for $j \in s_k$.
Of course, the $y_{kj}$ are known only for $j \in s_k$.
Estimation in both situations will be discussed.
We also assume that the finite population
and the observed sampling units can both be regarded as samples
from a common probability model, i.e., the sampling plan is uninformative.
The informative situation needs more careful treatment
(Guadarrama, Molina, and Rao 2016).

We are interested in predicting
finite-population parameter values under some model assumptions.
Most finite-population parameters of interest have the following algebraic form:
\begin{equation}
\label{MRpara}
H_k = N_k^{-1} \sum_{j=1}^{N_k} h(y_{kj})
\end{equation}
for some known function $h(\cdot)$.
When $h$ is chosen as $h(y) = y$, $H_k$ is the small area mean.
When $h(y) = \ind(y \leq t)$ for some real value $t$, where $\ind(\cdot)$
is an indicator function,
$H_k$ is the small area cumulative distribution function
$F_k(t)$ at $t$. The small area quantile function is the
inverse of $F_k(t)$.
We refer to Molina and Rao (2010) for
additional examples.

Under a probability model on the finite population,
the minimum variance unbiased
prediction (when feasible) of $H_k$ is given by
\[
E(H_k | \mbox{sampled information})
=
N_k^{-1} \sum_{j=1}^{N_k} E\{h(y_{kj}) | \mbox{sampled information}\}.
\]
If the resulting conditional expectation contains unknown model parameters,
the prediction will be constructed with the unknown parameters
replaced by suitable estimates. This leads to
the empirical best predictor(s) (EBP) of Molina and Rao (2010):
\begin{equation}
\label{EBP}
\hat H_k =
N_k^{-1} \big \{ \sum_{j \not \in s_k } \hat h_{kj}
+  \sum_{j \in s_k} h(y_{kj}) \big \}
\end{equation}
where $\hat h_{kj}$ is the predicted value of $h(y_{kj})$.

In applications, it can be difficult to identify $s_k$ from the
finite population. Hence, we may use its census version
\begin{equation}
\label{MRcensus}
\hat H_k^{\mbox{\sc c}} = N_k^{-1} \sum_{j =1}^{N_k} \hat h_{kj}.
\end{equation}
The EBP works well, but establishing its optimality
can be a challenging task.

Once a concrete model is given, the abstract EBP
becomes a practical solution.
On the model front, the nested-error (unit level) regression model
(NER) of Battese, Harter, and Fuller (1988) is  widely adopted.
Under this model,
\ba
\label{ner}
y_{kj} =  \bx_{kj}^{\tau} \bbeta + \nu_k +  \varepsilon_{kj},
\ea
where $\nu_k$ denotes an area-specific random effect and
$\varepsilon_{kj}$ is random error.
The homogeneous NER model assumptions include
\( \nu_k \sim N(0, \sigma_v^2) \),
\( \varepsilon_{kj} \sim N(0, \sigma_e^2) \),
and they are independent of each other and the auxiliary variable $\bx_{kj}$.
Relaxing the homogeneity to a more flexible variance
structure leads to the heterogeneous NER (HNER) of Jiang and Nguyen (2012).
Relaxing the normality of the error distribution
to a skewed normal distribution is discussed by
Diallo and Rao (2016).
Recent extensions include replacing $\bx_{kj}^{\tau} \bbeta$
with a spline (Opsomer et al. 2008; Ranalli,  Breidt, and Opsomer 2016).
One may also transform $y_{kj}$
to make the normality assumption more appropriate (Molina and Rao 2010).
%% Wait for further information (Rao, 2017) .

Under NER or HNER models, the regression coefficient $\bbeta$
is common across the small areas.
Samples from all the areas contain its information.
When the overall sample size $n= \sum_{k=0}^m n_k$ is large,
a high precision estimator $\hat \bbeta$ is possible.
Given the population means $\bar{\bX}_{k}$,
we get an indirect estimator $\hat{\bar{\bY}}_{k} = \bar{\bX}_{k}^\tau \hat \bbeta$.
It may be optimally combined with the regression estimator
$\bar y_k + (\bar{\bX}_k - \bar{\bx}_k) \hat{\bbeta}$
in obvious notation to get the so-called BLUP of small area mean $\bar Y_k$.
The linear combination coefficient depends on whether the
NER or HNER model is assumed (Jiang and Lahiri 2006;
Jiang and Nguyen 2012).

Another general approach is via calibration or generalized regression
(Estevao and S\'{a}rndal 2006; Pfeffermann 2013).
Suppose $\hat h_{kj}$ predicting $h(y_{kj})$ is available for all the units in the finite population.
A calibration predictor of $H_k$ is given by
\begin{equation}
\label{Calibrate}
\hat H_k =
N_k^{-1}  \sum_{j=1}^{N_k} \hat h_{kj}
+  N_k^{-1} \sum_{j \in s_k} w_{kj} \{ h(y_{kj}) - \hat h_{kj}\}
\end{equation}
where the $w_{kj}$ are design weights to reduce the risk of
bias caused by informative sampling plans, and
$s_k$ denotes the sample of units selected from area $k$.
Under a simple random sample without replacement plan or if the sampling
plan is non-informative, we may use $w_{kj} = N_k/n_k$.
Specifically, under linear models such as NER, $\hat h_{kj}$
is generally chosen to be $\bx_{kj}^\tau \hat \bbeta$ leading to the
generalized regression estimator (GREG); see
Pfeffermann (2013). In this case, the calibration
estimator improves the efficiency
of sample mean $\bar y_k$ by calibrating the difference
between $\bar \bx_k$ and $\bar \bX_k$.
In nonlinear situations, this approach needs census information
on $x$ and calibrates only the difference between two averages:
$N_k^{-1}  \sum_{j=1}^{N_k} \hat h_{kj}$ and $N_k^{-1}  \sum_{j\in s_k} w_{kj} \hat h_{kj}$.
Hence, it is not a good choice for the estimation of quantiles.

Another choice of $\hat h_{kj}$ is via the M-quantile (Breckling and Chambers 1988).
A regression quantile relates the response variable $Y$
and some covariate $\bx$ through the equation
\[
P( Y \leq \bx^\tau \bbeta_q|\bX = \bx) = q
\]
for each $q \in (0, 1)$ and a $q$-dependent $\bbeta_q$;
see Koenker and Bassett (1978).
Let $\rho_q(t) = q \ind (t < 0) + (1-q) \ind (t > 0)$.
Then $\bbeta_q$ is also a solution to
\[
\min \e \{\rho_q(Y - \bX^\tau \bbeta)|\bX\}.
\]
By this statement, we have implicitly assumed that the
solution to the above equation in $\bbeta$ does not depend
on the value of $\bX$.
When the model is valid, $\bx^\tau \bbeta_q$ is the $q$th quantile
of the conditional distribution of $Y$ given $\bX = \bx$.
Clearly, $\bX^\tau \bbeta_q$ is a robust description
of the conditional distribution of $Y$.
Breckling and Chambers (1988) propose the
use of a generic $\rho_q(\cdot)$ function (say $\psi$)
and call the resulting $\bX^\tau \bbeta_q$ the M-quantile.

In the context of small area estimation, let $\hat \bbeta(q)= \hat  \bbeta_q$
be the fitted M-quantile given $q \in (0, 1)$. Note that it depends on $q$.
For each unit $k, j$ in the sample, one may find a $q$
value such that
\[
y_{kj} = \bx_{kj}^\tau \hat \bbeta(q).
\]
An approximation may be used when an exact solution does not exist.
Denote the solution as $q_{kj}$.
Chambers and Tzavidis (2006) suggest that
the average $q_{k\cdot} = n_k^{-1} \sum_{j=1}^{n_k} q_{kj}$
reflects the general quantile information of area $k$.
This leads to $\hat y_{kj} = \bx_{kj}^\tau \hat \bbeta(q_{k\cdot})$,
the predicted area-specific cumulative distribution function
\[
\hat F_k(t)
=
{N_k^{-1} } \big [ \sum_{j \in s_k} \ind (y_{kj} \leq t)
+
\sum_{j \not \in s_k} \ind \{\bx_{kj}^\tau \hat \bbeta(q_{k\cdot})\leq t\} \big ],
\]
and the resulting quantile predictions.

As pointed out by Tzavidis and Chambers (2005) and Tzavidis et al.~(2008),
from $F_k(t)$ to $\hat F_k(t)$ the difference between
$\ind(\bx_{kj}^\tau \bbeta + \epsilon_{kj} \leq t )$ and $\ind(\bx_{kj}^\tau \bbeta \leq t )$
is ignored, which leads to a nondiminishing error even when $n_{k} \to \infty$.
To overcome this pitfall, a new estimator/predictor following the approach
of Chambers and Dunstan (1986) is proposed.
Let $\hat \epsilon_{kj} = y_{kj} - \hat y_{kj}$ be the M-quantile
residuals for $j \in s_k$ over $k=0, 1, \ldots, m$,
where $\hat y_{kj} = \bx_{kj}^\tau \hat \bbeta(q_{k\cdot})$.
For each small area, construct an empirical distribution
\[
\hat G_k(t) = n_k^{-1} \sum_{j \in s_k} \ind( \hat \epsilon_{kj} \leq t).
\]
The revised estimate of $F_k$ (Tzavidis et al.~2008) can be written
\begin{equation}
\label{Mquantile}
\hat F^{\mbox{\sc mq}}_k (t) =
N_k^{-1} \big \{
\sum_{j \in s_k} \ind (y_{kj} \leq t)
+
\sum_{j \not \in s_k} \hat G_k (t - \hat y_{kj}) \big \}.
\end{equation}
Note that we have written this estimator in the form of the EBP of
{ Molina and Rao (2010)}. The approach may also
be made outlier-robust {(Chambers et al., 2011)}.

This paper provides a new approach to the prediction of small area quantiles.

\section{The proposed approach}

We assume the basic NER model structure \eqref{ner}
but allow a generic $G_k$ for the distribution of $\varepsilon$,
the expectation of which is zero.
Hence,
\[
\e\{ \ind (y_{kj} \leq y)| \nu_k, x_{kj}\}
= G_k (y - \nu_k - \bx_{kj}^\tau \bbeta).
\]
Based on a random sample $s_k$ and when feasible,
we predict $F_k(y)$ by
\begin{equation}
\label{delta}
\tilde F_k(y) =
n_k^{-1} \sum_{j \in s_k}
G_k (y - \nu_k - \bx_{kj}^\tau \bbeta - \delta_k),
\end{equation}
with $\delta_k$ chosen to permit the shrinkage
effect via random effect considerations.

When census information on $x$ is available,
we follow the principle of EBP (Molina and Rao 2010)
to predict $F_k(y)$ by
\[
\tilde F^{\mbox{\sc eb}1}_k(y)
= N_k^{-1} \big \{
\sum_{j\not \in s_k} G_k (y - \nu_k - \bx_{kj}^\tau \bbeta)
+
\sum_{j \in s_k}  \ind(y_{kj} \leq y) \big \}.
\]
If the identification of $s_k$ is difficult, then the following predictor
is just as effective:
\[
\tilde F^{\mbox{\sc eb}2}_k(y)
= N_k^{-1}  \sum_{j=1}^{N_k}  G_k (y - \nu_k - \bx_{kj}^\tau \bbeta).
\]

Since $\nu_k$, $\bbeta$, and $G_k$ are not known in applications in general,
it is common practice to replace them in the above expressions
by their predictions/estimates.
This leads to a variety of predictors.
Let $\hat F_k(y)$ be a generic predictor of the small area distribution.
The corresponding small area quantiles predictor will be defined as
\ba
\label{ELquantile}
\hat \xi_k = \hat \xi_{k, \alpha} = \inf\{ y: \hat F_k(y)\geq \alpha\}
\ea
for any $\alpha \in (0, 1)$.
The remaining tasks are to choose $\delta_k$,
estimate $G_k$, and predict the other quantities.

\subsection{Estimation under the NER model}

Under NER, we can estimate the unknown parameters via the maximum likelihood.
Let $\tilde \sigma^2$, $\tilde \sigma_v^2$, and $\tilde \bbeta$
be the MLEs.
An established small area mean estimate is the empirical BLUP (EBLUP)
given by
\ba
\label{eblup}
\tilde{\bar{Y}}_k
=
  \bar{\bX}_{k}^{\tau}  \tilde \bbeta
+
\tilde \gamma_k ( \bar{y}_{k } -  \bar{\bx}_{k}^{\tau}  \tilde \bbeta)
=
 \bar{\bX}_{k}^{\tau}  \tilde \bbeta
+
\tilde \gamma_k  \tilde \nu_k
\ea
where
$\tilde \gamma_k = n_k \tilde \sigma_v^2/(\tilde \sigma^2+n_k \tilde \sigma_v^2)$
and $\tilde \nu_k = \bar{y}_{k } -  \bar{\bx}_{k}^{\tau}  \tilde \bbeta$.
Note that the EBLUP has shrunk $\tilde v_k$ toward zero
by modeling $v_k$ as a random effect.
Let $\delta_k = \tilde{\bar{Y}}_k - \bar{\bx}_k^\tau \tilde \bbeta$
in \eqref{delta};
we then get a predictor as
\ba
\label{cdf-ner}
\hat F^{\mbox{\sc ner}}_k(y)
=
\frac{1}{n_k} \sum_{j=1}^{n_k}
\Phi
\left(
\{ y - (\bx_{kj}- \bar \bx_k)^{\tau} \tilde \bbeta - \tilde{\bar{Y}}_k \}/ \tilde \sigma_e
\right).
\ea
The mean of the distribution $\hat F^{\mbox{\sc ner}}_k(y)$ is
exactly $\tilde{\bar{Y}}_k$ because of the choice of $\delta_k$.

When the census $x$ information is available,
the EBP versions of $\hat F^{\mbox{\sc ner}}_k(y)$ are given by
\ba
\label{EBP-ner1}
\hat F^{\mbox{\sc eb}1}_k(y)
= N_k^{-1} \big \{
\sum_{j\not \in s_k}
\Phi
(
\{ y - \tilde \nu_k - \bx_{kj}^{\tau} \tilde \bbeta\}/ \tilde \sigma_e
)
+
\sum_{j \in s_k} \ind(y_{kj} \leq y) \big \}
\ea
and
\ba
\label{EBP-ner2}
\hat F^{\mbox{\sc eb}2}_k(y)
= N_k^{-1}
\sum_{j=1}^{N_k}
\Phi
( \{ y - \tilde \nu_k - \bx_{kj}^{\tau} \tilde \bbeta\}/ \tilde \sigma_e  ).
\ea
%The mean of the distribution $\hat F^{\mbox{\sc eb}2}_k(y)$
%is exactly the desirable
%$\tilde \nu_k + \bar X_k^\tau \tilde \bbeta
%=
%\bar y_k  + (\bar X_k - \bar x_k)^\tau \tilde \bbeta$.

\subsection{Estimation under DRM}

As pointed out by Diallo and Rao (2016), the normality assumption
on the error distribution of $\varepsilon$ can have a marked influence
on the estimation of $F_k$. To alleviate this concern, a skewed
normal distribution can be used. In this paper, we adopt a
semiparametric density ratio model (DRM) for $G_k$ (Anderson 1979):
\ba
\label{DRM}
\log \{dG_k(t)/dG_0(t)\} = \btheta_k^{\tau} \bq(t),
\ea
with a prespecified $d_2$-variate function $\bq(t)$
and area-specific tilting parameter $\btheta_k$.
We require the first element of $\bq(t)$ to be one so that
the first element of $\btheta_k$ is a normalization parameter.
The baseline distribution $G_0(t)$ is left unspecified, and
there many potential choices of $\bq(t)$.
%%%%
The nonparametric $G_0$ has abundant flexibility while
the parametric tilting factor $\btheta_k^{\tau} \bq(t)$ enables
effective ``strength borrowing'' between small areas.
Note also that any $G_j$, not just $G_0$,
may be regarded as a baseline distribution because
\ba
\log \{dG_k(t)/dG_j(t)\} = (\btheta_k- \btheta_j)^{\tau} \bq(t).
\ea
DRM is flexible, as testified by its inclusion of the normal, Gamma,
and many other distribution families.
Under this model assumption, we look for an estimate of $G_k$.

\vspace{1ex}
\noindent
{\bf Estimating $G_k$ under DRM.}
\label{sec3.1}

Consider an artificial situation where we have $m+1$ samples
$\{\varepsilon_{kj}: j=1, 2, \ldots, n_k; k=0, \ldots, m\}$ from a DRM.
Following Owen (1988, 2001) or Qin and Lawless (1994),
we confine the form of the candidate $G_0$ to
$G_0 (t) = \sum_{k, j} p_{kj} \ind (\varepsilon_{kj} \leq t)$,
and the summation  $\sum_{k, j}$ is short for
$\sum_{k=0}^m \sum_{j=1}^{n_k}$.
The support of $G_0$ includes
all $\varepsilon_{kj}$, not just those with $k=0$.
This is part of the strength-borrowing strategy.
In this setting, $p_{kj} = dG_0(\varepsilon_{kj})$ and
\(
dG_k(\varepsilon_{ij})
=   p_{ij}  \exp\{ \btheta_k^{\tau} \bq(\varepsilon_{ij}) \}, \; k=0, 1, \ldots, m,
\)
where $\btheta_k$ are $d_2$-variate unknown parameters,
and
\ba
\label{gkt}
G_k(t)
= \sum_{i, j} p_{ij} \exp\{ \btheta_k^{\tau} \bq(\varepsilon_{ij}) \} \ind (\varepsilon_{ij} \leq t).
\ea
Clearly, $\btheta_0 = 0$ when $G_0$ is chosen as the baseline.
Because $\varepsilon_{kj}$ follows $G_k(t)$, it contributes
to the likelihood only through $dG_k(\varepsilon_{kj})$.
This leads to the empirical likelihood (EL):
\bas
L_n(G_0, G_1, \ldots, G_m)
=
\prod_{k,j} dG_k(\varepsilon_{kj})
=
\big ( \prod_{k,j} p_{kj} \big )
\cdot \exp \big [
\sum_{k,j} \{  \btheta_k^{\tau} \bq(\varepsilon_{kj}) \}
 \big ]
\eas
where the $p_{kj}$'s satisfy
$p_{kj}\geq 0$ and
for all $k=0, 1, \ldots, m$,
\ba
\label{constr}
\sum_{i,j} p_{ij}  \exp\{\btheta_k^{\tau} \bq(\varepsilon_{ij})\}  = 1.
\ea
%We use $i, j$ as dummy indexes in the above summation because $k$ is
%reserved as the index of the $k$th small area.

Let   $\btheta^\tau = (\btheta^\tau_1, \ldots, \btheta_m^\tau)$.
Maximizing  the empirical log-likelihood
\bas
\ell_n(\btheta, G_0)
=
  \sum_{k,j} p_{kj}
+
\sum_{k,j} \{  \btheta_k^{\tau} \bq(\varepsilon_{kj}) \}
\eas
with   respect to $G_0$ under constraints (\ref{constr}) results in the fitted probabilities
(Qin and Lawless 1994)
\ba
\label{pp}
\hat{p}_{kj} = n^{-1}
\{1+ \sum_{l=1}^m \lambda_l [\exp\{\btheta_l^{\tau} \bq(\varepsilon_{kj})\}- 1]  \}^{-1}
\ea
and the profile EL, up to an additive constant,
\[
\tilde{\ell}_n (\btheta)
=
- \sum_{k,j}
\log\{ 1+ \sum_{l=1}^m \lambda_l[\exp\{\btheta_l^{\tau} \bq(\varepsilon_{kj})\}- 1] \}
+ \sum_{k,j}\{ \btheta_k^{\tau} \bq(\varepsilon_{kj}) \}
\]
with $(\lambda_1, \lambda_2, ...,\lambda_m)$ being the solution to
\[
\sum_{i,j}
\frac{ \exp\{\btheta_k^{\tau} \bq(\varepsilon_{ij})\}- 1}{
1+\sum_{l=1}^m \lambda_l  [\exp\{\btheta_l^{\tau} \bq(\varepsilon_{ij})\}- 1] }
= 0
\]
for  $k=1, \ldots, m$.
The stationary points of $\tilde \ell_n (\btheta)$ coincide
with those of a dual form of the empirical log-likelihood function
(Keziou  and  Leoni-Aubin  2008)
\ba
\label{dual}
\breve{\ell}_n (\btheta)
=
-
\sum_{k, j}
\log \big [ \sum_{r=0}^m \rho_{r} \exp\{\btheta_r^{\tau} \bq(\varepsilon_{kj})\} \big ]
+
\sum_{k, j} \btheta_k^{\tau} \bq(\varepsilon_{kj}),
\ea
with $\rho_{r} = n_r/n$, $r=0, 1, \ldots, m$.

For point estimation, it is simpler to work with
$\breve{\ell}_n (\btheta)$, which is convex and free from constraints.
Once the values of $\varepsilon_{kj}$ are provided, it is relatively
simple to find its maximum point, which is the maximum EL
estimate of $\btheta$. We then use (\ref{pp}) to compute the
fitted values with $\lambda_l$  replaced by $\rho_l$.  We subsequently obtain
$\hat{G}_k$ and the other parameters of interest via the invariance principle.

This line of approach first appeared in Qin and Zhang (1997), Qin (1998),
Zhang (1997), and others. In particular, the properties of the quantile estimators
are discussed by Zhang (2000) and Chen and Liu (2013).
In the current application, we use $\hat \varepsilon_{kj}$,
given below in (\ref{residual}), for the computation.

\vspace{1ex}
\noindent
{\bf Parameter estimation with fitted residuals}

Suppose we have a sample $(y_{kj}, \bx_{kj})$ for $k=0, 1, \ldots, m$ and
$j=1, \ldots, n_k$ satisfying the NER with the error distribution from the DRM.
We first eliminate the random effect $\nu_k$ from the NER
by centralizing both sides of \eqref{ner},   which leads to
\bas
y_{kj} - \bar y = (\bx_{kj} - \bar \bx_k)^\tau \bbeta + \varepsilon_{kj} - \bar \varepsilon_k,
\eas
where $ {\bar \bx}_k$ and $\bar y_k$ are the sample means over small
area $k$.
The least squares estimator of $\bbeta$ under the centralized model is
\ba
\label{betahat}
\hat \bbeta
=
\{ \sum_{k,j}  (\bx_{kj} - {\bar \bx}_k)^\tau (\bx_{kj} - {\bar \bx}_k)  \}^{-1}
\{ \sum_{k,j}   (\bx_{kj} - {\bar \bx}_k)^\tau (y_{kj} - \bar y_k) \}.
\ea
%%% very good suggestion
The residuals of this fit are given by
\ba
\label{residual}
\hat{\varepsilon}_{kj}
=
y_{kj} - \bar y_k - (\bx_{kj} - \bar \bx_k)^\tau \hat \bbeta.
\ea
We then treat $\{\hat{\varepsilon}_{kj}: j=1, 2, \ldots, n_k\}$ as samples
from the DRM and apply the EL method of Section \ref{sec3.1}.

Let $\ell_n(\btheta)$ denote the log EL function (\ref{dual})
with $\varepsilon_{kj}$ replaced by $\hat \varepsilon_{kj}$.
We define the maximum EL estimator of $\btheta$ by
\(
\hat \btheta = \argmax \ell_n(\btheta)
\)
and accordingly  define the  estimators
\ba
\label{G_est}
\hat{G}_k(t)
=
\sum_{i,j}
\hat{p}_{ij} \exp\{\hat{\btheta}_k^{\tau} \bq(\hat{\varepsilon}_{ij})\}
\ind (\hat{\varepsilon}_{ij}<t)
\ea
with $\hat \btheta_0 = \0$ by convention and
\(
\hat{p}_{ij}
=
n^{-1}\{1+ \sum_{l=1}^m \rho_l
[\exp\{ { \hat{\btheta}_l^{\tau} } \bq(\hat{\varepsilon}_{ij})\}- 1]  \}^{-1}.
\)
Consequently, after targeting the small area mean estimate in
\eqref{cdf-ner}, we estimate $F_k(y)$ by
\ba
\label{cdf-EL}
\hat F^{\mbox{\sc el}}_k(y)
=
\frac{1}{n_k} \sum_{j=1}^{n_k}
\hat G_k
\left(
y - (\bx_{kj}- \bar \bx_k)^{\tau} \hat \bbeta - \tilde{\bar{Y}}_k
\right)
\ea
where $\tilde{\bar{Y}}_k$ is given in \eqref{eblup}.
When the census $x$ information is available, the EBP versions
are
\begin{equation}
\label{CensusEL-DRM-EBP1}
\hat F^{\mbox{\sc ebel}1}_k(y)
=
N_k^{-1} \big \{
 \sum_{j \not \in s_k} \hat G_k (y - \hat  \nu_k - \bx_{kj}^\tau \hat \bbeta)
+
 \sum_{j \in s_k} \ind(y_{kj} \leq y) \big \}
\end{equation}
where
\(
\hat{\nu}_k
=
\bar y_k - \bar{\bx}_{k}^\tau \hat \bbeta
\),
%is a natural prediction of $\nu_k$.
% %%% I start to wonder why we should not use shrunk \nu_k?
%When $s_k$ is hard to identify from the census $x$ values,
%we use
and
\begin{equation}
\label{CensusEL-DRM-EBP2}
\hat F^{\mbox{\sc ebel}2}_k(y)
=
N_k^{-1}
 \sum_{j=1}^{N_k} \hat G_k (y - \hat  \nu_k - \bx_{kj}^\tau \hat \bbeta).
\end{equation}
The quantiles are estimated accordingly.

\section{Variance/MSE estimation}
When an estimator is assembled in many steps,
its variance is often too complex to be analytically evaluated. Resampling
the variance estimation becomes a good choice (Molina and Rao 2010).
Based on whether or not census information is available and
whether the error distribution is regarded as $N(0, \sigma_e^2)$ under
the NER or $G_k$ under the DRM, we have four distinct small area
quantile estimators. We give a detailed description of a
resampling method for the case where census information is
available and the error distributions $G_k$ satisfy the DRM.
We then give a simple description of the changes needed
for the other three estimators.

Our resampling procedure is as follows:
\begin{enumerate}
\item
Under the NER model, obtain the maximum likelihood
estimates $\tilde \sigma_v^2$ and   $\tilde \sigma_e^2$,
and compute $\tilde{\bar Y}_k$.
\item
Calculate $\hat \bbeta$ and
 obtain $\hat \btheta_k$ and $\hat G_k$ as in \eqref{G_est} under DRM.
\item
\label{Bootstep}
For $b=1, \ldots, B$ over $k, j$ with $B$ large, generate
\[
\nu_k^{*(b)} \sim N(0, \tilde \sigma_v^2) ~\mbox{and} ~ e_{kj}^{*(b)} \sim \hat G_k.
\]
\item
Construct $B$ (conditionally) independent and identically
distributed (iid) bootstrap populations with
\[
y_{kj}^{*(b)}
=
\bx_{kj}^\tau \hat \bbeta + \nu_k^{*(b)} +  e_{kj}^{*(b)}
\]
for $j=1, \ldots, N_k$ and $k=0, 1, \ldots, m$.
\item
For each bootstrap sample, compute
\[
F^{*(b)} _k(t) = N_k^{-1} \sum_{j=1}^{N_k} \ind ( y_{kj}^{*(b)}  \leq t)
\]
and the corresponding
$\hat{F}^{*(b)} _k(t)$ as in \eqref{CensusEL-DRM-EBP2}.
\item
\label{mseStep}
For any parameter that can be written in the form of $H(F_k)$,
compute the bootstrap mean square error estimator of
MSE($H(\hat F_k)$) via
\begin{equation}
\label{mse}
\mbox{mse}(H(\hat F_k))
=
\frac{1}{B} \sum_{b=1}^B \{ H(\hat{F}_k^{*(b)}) - H(F_k^{*(b)})\}^2.
\end{equation}
\end{enumerate}
Sampling from $\hat G_k$ can easily be  done with existing
R functions because it is a discrete distribution on
$\hat \epsilon_{ij}$ with probabilities
$\hat p_{ij} \exp \{ \hat \btheta_k \bq(\hat \varepsilon_{ij})\}$.
Note that the support is over all the fitted residuals, not just those in small area $k$.

Under the NER, we replace $\hat G_k$ in Step \ref{Bootstep} by $N(0, \tilde \sigma_e^2)$.
Under the DRM without census information, we generate $\epsilon_{kj}$ in
Step \ref{Bootstep} only for $j \in s_k$ and in Step \ref{mseStep}
we use the sample variance of  $H(\hat{F}_k^{*(b)}) - H(F_k^{*(b)})$
instead of the squared average.

%%% Yukun: Am I right?

%%%  To Prof. Chen:
%%%       Yes.  You are right.

\section{Asymptotic properties}
For each $k$,
the covariates $\{  \bx_{kj},    j=1, 2, \ldots, n_k\}$ are iid
with finite mean and nonsingular and finite covariance matrix $\bV_k$;
the error terms $\{\varepsilon_{kj}: j=1, 2, \cdots, n_k\}$
are iid samples, independent of the covariates, with
conditional variance $\sigma^2_k$.
The pure residuals $\varepsilon_{kj}$ form $m+1$ samples
from populations with the distribution function $G_k$
satisfying (\ref{DRM}).
Let the total sample size $n = \sum_k n_k \to \infty$, and
assume $\rho_k = n_k/n$ remains a constant (or within an $n^{-1}$ range)
as $n$ increases. Let  $\hat \bbeta$ and $\hat \btheta$
be defined by (\ref{betahat}) and the subsequent steps.

\begin{theorem}
\label{normality-lse}
Assume the general setting presented in this subsection.
Let  $\bV_x = \sum_{k=0}^m \rho_k \bV_k$.
As $n \rightarrow \infty$,  we have
\(
\sqrt{n}(\hat \bbeta - \bbeta) \overset{d}{\longrightarrow} N(0, \bSigma_{\bbeta} ),
\)
where $\overset{d}{\longrightarrow}$ denotes convergence in distribution and
\(
\bSigma_{\bbeta}
= \bV_x^{-1}
(  \sum_{k}  \rho_k \bV_k \sigma_k^2)
\bV_x^{-1}
\).
\end{theorem}

%The result on asymptotic normality of the maximum EL estimator of $\btheta$
%based on iid $\varepsilon_{kj}$'s can be found in Chen and Liu (2013).
%We will note that $\hat \varepsilon_{kj} \approx \varepsilon_{kj}$ uniformly.
%It is hence natural to expect that $\hat \btheta$,
%which are based on $\hat \varepsilon_{kj}$'s,
%are also asymptotically normal.

For ease of exposition of the next theorem, we introduce some notation.
For $k=0, 1, \ldots, m$, let
\[
h(x; \btheta)
=
\sum_{k=0}^m  \rho_{k} \exp\{ \btheta_k^{\tau} \bq(x)\};
~~~
h_k(x; \btheta)
=
\rho_{k} \exp\{ \btheta_k^{\tau} \bq(x)\}/h(x; \btheta).
\]
Clearly, $0 < h_k < 1$ for all $k$.
Let
$\bh (x; \btheta)  = \{ h_0(x; \btheta),  \ldots, h_m(x; \btheta)\}^\tau$
and define an $(m+1) \times (m+1)$ matrix
\[
\bH(x; \btheta)
= \diag \{  \bh (x; \btheta)\}
- \bh (x; \btheta) \bh^\tau (x; \btheta).
\]

%\overset{\sqbullet}{x}
We will use $h(x; \overset{\sqbullet}{x})$ and
$h(x; \overset{\sqbullet}{\btheta})$ for
the partial derivatives of $h$ with respect to $x$ and $\btheta$ respectively.
When $\btheta = \btheta^*$, the true value of $\btheta$,
we may drop $\btheta^*$ in $h(x; \btheta^*)$ and denote it as $h(x)$.
Lastly, we use  $d\bar G(x)$ for $h(x; \btheta^*) dG_0(x)$ in the integrations.
%%% these seem not needed unless in appendix

\begin{theorem}
\label{normality-theta}
Assume the conditions of Theorem \ref{normality-lse}.
Furthermore, assume the population distributions $G_k$
satisfy the DRM (\ref{DRM}) with the true
parameter value $\btheta^*$, and
$\int h(t; \btheta) d G_0 < \infty$ in a neighborhood of $\btheta^*$.
Assume the components of  $\bq(t)$ are linearly independent
with the first element being one, twice differentiable,
and that there exist a  function $\psi(t)$ and $c_0>0$
such that
\ba
\label{moment}
\sup_{t: |t-u| \leq c_0}
\{ \| \bq(t) \|\|{\bq}( \overset{ \sqbullet \sqbullet}{t} ) \|  +
\| \bq(t) \|\|{\bq}(\overset{ \sqbullet}{t}  ) \|^2\}
\leq \psi(u)
\ea
for all $u$ with $\int \psi(u) d\bar G(u)< \infty$.
Then as $n$ goes to infinity,
$\sqrt{n}(\hat \btheta - \btheta^*) \converge N(\0, \bOmega)$ where
$\bOmega$ is  given in {(A.10) in the supplementary material.}
\end{theorem}

The assumption that
$\int h(t; \btheta) dG_0(t) < \infty$ in a neighborhood
of $\btheta^*$ implies the existence of the
moment generating function of $\bq(t)$
and therefore all its finite moments.
%Note that our log EL is
%\ba
%\label{emplik}
%\ell_n (\btheta)
%=
%\sum_{k, j}
%[ \btheta_k^{\tau} \bq(\hat \varepsilon_{kj}) -
%\log\{ h(\hat \varepsilon_{kj}; \btheta)\} ].
%\ea
%As the first component of $\bq(t)$ is equal to one,
%the condition in (\ref{moment}) implies
% \ba
%\label{moment2}
%\sup_{t: |t-u|\leq c_0}
% \|  \bq(\overset{\sqbullet\sqbullet}{t}) \|  +
% \|  \bq(\overset{\sqbullet}{t}) \|^2 \leq \psi(u) \;
%\mbox{for all }\; u.
%\ea
%Therefore for all $\{k,j\}$, when
%we approximate
%$\hat \btheta_k^{\tau} \bq(\hat \varepsilon_{kj}) -
%\log\{ h(\hat \varepsilon_{kj}; \hat \btheta)\}$
%by their first-order Taylor expansions at
%$(\varepsilon_{kj};  \btheta^*)$,
%the summation of the remainders is of order $o_p(1)$.
%Combining with the proof of Theorem 2.1 of Chen and Liu (2013),
%this immediately leads to the root-$n$ consistency of $\hat \btheta$.
%This observation alleviates the burden of proof.

%% If q(y) = log y, this condition is a bit problematic.
%%%  Gamma distribution with low degrees of freedom is an issue.
%% Think about it.

We next examine the asymptotic
properties of the proposed small area quantile estimators,  which we call EL quantiles for short.

\begin{theorem}
\label{normality-xi}
Assume the conditions of Theorem \ref{normality-theta}.
Suppose in addition that the $G_k(t)$ have smooth and bounded density functions,
and $F_k(y)$ has positive density at $\xi_k$.
Then the EL quantile (\ref{ELquantile}) based on
\eqref{cdf-EL} is root-$n$ consistent.
That is, $\hat \xi_k - \xi_k = O_p(n^{-1/2})$.
\end{theorem}

\section{Simulation study}

In this section, we investigate the
performance of various small area quantile estimators
and their variance estimates.
In the simulation, we examine the
$5\%$,  $25\%$, $50\%$, $75\%$, and $95\%$
small area quantile estimations.

\subsection{Simulation settings}

The first task of the simulation is to create finite populations.
We consider the following model for the general structure of
the population:
\ba
y_{kj} &=& \bx_{kj}^{\tau}\bbeta+ \nu_k + \varepsilon_{kj}.
	\label{model1}
\ea
For authenticity, we use real survey data
as a blueprint to design
the following simulation populations:
\begin{enumerate}
\item
For each $k=0, 1, \ldots 19$,  generate   $N_k = 1000$
three-dimensional
$\bx_{kj}=(x_{kj1}, x_{kj2}, x_{kj3})$ values,
where
$x_{kj1} \sim U(0, 50)$, $x_{kj2}=50z_{kj}$,
$z_{kj} \sim Beta(0.6, 0.6)$,
and conditional
$x_{kj3}|z_{kj} \sim \mbox{Binom}(12,  \; 0.6 + 0.1x_{kj2} )$.

\item
Let $\bbeta_0^\tau = (0.019, 0.022, 0.074)$.

\item
Generate $\nu_k$ from $N(8, 1)$.
\end{enumerate}

For the error distribution,
we generate $\varepsilon_{kj}$  from
\begin{itemize}
\item[]
(i) $N(0, \sigma_e^2)$ with $\sigma_e^2 = 2$;
\item[]
(ii) normal mixture
$0.5 N(- \mu_k/6,1) + 0.5 N(\mu_k/6, 1)$;
%% error variance is 1 + 2*0.5*(mu/6)^2
\item[]
(iii)
normal mixture
$0.1 N(- \mu_k/2,1) + 0.9 N(\mu_k/18, 1)$;
%% error variance is 1 + 0.1*(mu/2)^2+ .9*(mu/18)^2
\item[]
(iv)
normal mixture
$0.9 N(- \mu_k/18,1) + 0.1 N(\mu_k/2, 1)$.
\end{itemize}
A single error distribution chosen from the above is applied
to all the small areas.
Each of them either
(i) satisfies the NER model assumption;
(ii) is non-normal but symmetric;
(iii) is skewed to the right;
or (iv) is skewed to the left.

We generate $\mu_k$ in (ii)--(iv) from the uniform distribution on
the interval $[4.5, 6]$ to determine the impact of
mildly different error distributions in different small areas.

\subsection{Predictors in the simulation}

We study the performance of seven representative quantile predictors.
Their corresponding area population distribution predictors
are as follows.

\begin{enumerate}
\item
Direct Predictor (DIR): we compute the sample quantiles
for small area $k$ based on the sampled response values $y_{kj}$.

\item
The NER-based predictor (NER): This predictor
$\hat F^{\mbox{\sc ner}}_k(y)$
is defined in \eqref{cdf-ner} assuming that the error
distribution is normal. It uses only sampled $x$ information and
the known population mean $\bar{\bX}_k$ for each small area.

\item
The EL-based predictor (EL): This predictor $\hat F^{\mbox{\sc el}}_k(y)$
is defined in \eqref{cdf-EL}.
It uses only sampled $x$ information and
the known subpopulation mean $\bar \bX_k$ of each small area.

\item
The NER-based census predictor (EB):
This predictor
$\hat F^{\mbox{\sc eb}2}_k(y)$ is defined in
\eqref{EBP-ner2} assuming that the error distribution is normal.
The other predictor $\hat F^{\mbox{\sc eb}1}_k(y)$
leads to nearly identical performance for the quantile estimation.
To save space, $\hat F^{\mbox{\sc eb}1}_k(y)$ is not included in
the simulation.

\item
The proposed census predictor $\hat F^{\mbox{\sc ebel}2}_k(y)$ (EBEL):
This estimator is given in \eqref{CensusEL-DRM-EBP2}.
It is an analog of $\hat F^{\mbox{\sc eb}2}_k(y)$ except for
using an EL-DRM-based estimate of the error distribution in the
linear-model setting.

\item
The EBP of Molina and Rao (MR): This is the predictor specified in
\eqref{EBP} under the NER model.
Additional implementation details are given below.
The conditional distribution of $y_{kj}$ given sample $s_k$
can be expressed as
\begin{equation}
\label{MR1}
y_{kj|s} = \mu_{kj|s} + u_k+ \epsilon_{kj}
\end{equation}
with the conditional mean
\(
\mu_{kj|s} = \bx_{kj} \bbeta + \gamma_k(\bar y_k - \bx_k^\tau \bbeta)
\),
area-specific conditional random effect
$u_k \sim N(0, (1-\gamma_i) \sigma_v^2)$,
and conditional residual error $\epsilon_{kj} \sim N(0, \sigma_e^2)$.
The nonrandom constants and unknown parameter values of
$\gamma_k$, $\sigma_v^2$, $\sigma_e^2$ are replaced by
their estimated values (MLE in our simulation) in the computation.
%% Yukun: double check. MLS replaced by MLE.
With this preparation, we generate $y_{kj}^{(\ell)}$  for each $j \not \in s_k$
according to \eqref{MR1} for $\ell = 1, 2, \ldots, L=100$.
The corresponding empirical distribution
\[
\hat F_{kj|s}(t) = L^{-1} \sum_{\ell = 1}^{L} \ind( y_{kj}^{(\ell)} \leq t)
\]
is used to form the predictor
\begin{equation}
\label{MRp}
\hat F_k^{\mbox{\sc mr}}(t)
=
N_k^{-1} \big \{ \sum_{j \not \in s_k }\hat F_{kj|s}(t)
+  \sum_{j \in s_k} \ind(y_{kj} \leq t)  \big \}
\end{equation}
and the corresponding quantile predictions.

\item
The M-quantile predictor (MQ): this predictor is specified in \eqref{Mquantile},
and it is also a census predictor.
Additional implementation details must be specified.
We use
\[
\psi(u; q) =q \ind (u > 0) - (1-q) \ind (u \leq 0)
\]
for $q \in (0, 1)$. For each $q = \{1, \ldots, 199\}/200$ and small area $k$,
we search for a solution in $\bbeta$ to
\[
\sum_{j \in s_k} \psi(y_{kj} - \bx_{kj}^\tau \bbeta; q) \bx_{kj}^\tau  = 0.
\]
Denote the solution as $\hat \bbeta_k(q)$.
For each $y_{kj} \in s_k$, we find a q-value in $\{1, \ldots, 199\}/200$
that minimizes $|y_{kj} - \bx_{kj} \hat \bbeta_k(q)|$, and this gives us $q_{kj}$.
The other numerical details have been given earlier.

\end{enumerate}

We do not include the method of Elbers, Lanjouw, and Lanjouw (2003)
because it is not designed specifically for
small area quantile estimation, and its properties have been
well investigated by Molina and Rao (2010).
We exclude from the simulation some of the other predictors discussed in this paper.
Preliminary experiments indicated that
they did not outperform the predictors that we have included.

One must specify $\bq(t)$ in the EL-DRM-based estimators (EL2, EBEL2).
There are many reasonable candidates, and after some
experiments, we settled on $\bq(t) = (1, \mbox{sign-root}(t))^{\tau}$.
It is not uniformly the best choice.
To reduce the amount of computation, we included only
this choice in our simulation.
In applications, mild model violation is unavoidable.
This choice is motivated by its overall performance
in terms of ``model robustness.''

The seven predictors listed above form two groups: the first
group does not use census $x$ information and the second group does.
Their performance will be judged in light of this difference.

\subsection{Performance measures}

Let ${\hat{\xi}}_{k}^{(j)}$ and ${{\xi}}_{k}^{(j)}$denote a generic quantile
estimate in the $j$th repetition and the corresponding population quantile.
We report the average mean squared error (\amse), defined to be:
\[
\amse
=  \{N(m+1)\}^{-1}
\sum_{k=0}^m  \sum_{j=1}^{N} ( {\hat{\xi}_{k}^{(j)} } - \xi_k^{(j)} )^2.
\]
This combines the loss of precision due to bias
and variation; it is a convenient metric of the performance of
different estimation methods. We find that using both variance and
bias does not lead to more detailed performance information but makes
the judgement burdensome.

%When the assumed model does not match the population,
%bias can be the main cause of the inflated \amse.
%For this reason, the average absolute biases is informative:
%\[
%\abias
%= (m+1)^{-1}
%\sum_{k=0}^m \big | N^{-1} \sum_{j=1}^{N} \{\hat{\xi}_{k}^{(j)} - \xi_k^{(j)}\} \big |.
%\]

\subsection{Simulation results}

We generate a new finite population for each simulation replication.
The small area population quantiles therefore vary
from replication to replication, which is necessary for assessing
the performance of the model-based methods.

We provide simulated \amse\ values of all the methods
for the populations generated with
$\bbeta = 1.0 \bbeta_0, 1.25 \bbeta_0$, and $1.5 \bbeta_0$.
These choices set the signal-to-noise ratios to around
30\%, 50\%, and 70\%, allowing us to determine the impact of this ratio
on the performance of the methods.
We choose two sample sizes: $n_k = 30, 50$
corresponding to the total sample size $n=600, 1000$ respectively.

Because the resampling method involves considerable
computation, the \amse\ estimates are calculated only for
$\bbeta = 1.5 \bbeta_0$ in two cases:
$n=600, n_k=30$ with $B = 100$ and 1000 repetitions;
$n=1000, n_k = 50$ with $B = 100$ and 500 repetitions.
To ease the computational burden, the resampling
is limited to DIR, EL, MR, and EBEL; the other methods clearly
have inferior performance in terms of \amse.
%We report the average ratios of estimated \amse\ and the
%simulated \amse\ cross the small areas. The \amse\ works
%best when this ratio is close to one.
We report the averages of the ratios of the estimated MSEs and the
simulated  MSEs  across all the small areas except
those with the largest two and smallest two simulated MSEs.
The closer the ratio to one,  the better the method.

Table \ref{table-amse-nk=30} presents the \amse\ values of the seven estimators
when the data are generated from model \eqref{model1} with
$\bbeta = 1.5 \bbeta_0$, $n=600, n_k=30$, and 1000 repetitions.
The ratios of the resampling estimated and simulated AMSEs
are given in Table \ref{table-ratio-nk=30}.
We summarize the results as follows:

\ben
\item
Under Scenarios (i) and (ii), where the error distributions
are normal or close to normal, NER and MR
are the winners, with EB the runner-up, and EL and EBEL performing nearly
as well. These methods have small and ignorable biases.

\item
Under Scenarios (iii) and (iv), where the violation of normality is
from moderate to severe,
EL and EBEL are clearly the winners.
They have much smaller AMSEs than the other methods,
particularly for the 5\% and 95\% quantiles.

\item
EL has surprisingly good performance,
although it does not use census information.

\item
The bootstrap MSE estimates work well for the DIR
quantile estimators in all scenarios, implying that the
resampling procedure is appropriate in general.

The bootstrap MSE estimates have
satisfactory precision for EL and EBEL in general,
but they mildly under-estimate those of EL for the 5\% quantile in
Scenario (iii) and the 95\% quantile in Scenario (iv).

The bootstrap MSE estimates work well for MR in Scenarios (i) and (ii)
but are less satisfactory in Scenarios (iii) and (iv), where the error distributions are non-normal.
This is understandable because
the version of MR used in our simulation
is based on the normality assumption.
This problem should disappear when the model assumptions
and the resampling procedure are in line.
\een

The top portions of the plots in Figures \ref{fig-s1} and \ref{fig-s2} depict
the area-specific MSEs of NER, EL, MR, EB, and EBEL.
DIR and MQ are not included because their MSEs are much larger;
including them masks the differences between the other methods.
The lower portions of the plots give the ratios of the estimated
and simulated AMSEs of EL, MR, and EBEL.
The ratios of the other methods are not included because they do not perform well.
The five plots in the left column are for Scenario (i),
and these in the right column are for Scenario (iv).
The results for Scenarios (ii) and (iii) are between those for (i) and (iv)
and are not shown.
 The plots provide quick visual summaries of the performance.
%%%   To Professor Chen:   This sentence does not read smooth.
%%%        However, I don't know how to correct it.
%% Should be okay Leave it to Julie.

There are six combinations of the sample sizes and signal-to-noise ratios.
We have presented just one combination here. To save
space, we include the results for the other five combinations in
the supplementary file.

\subsection{Illustration}

Finite populations created based on statistical models are inevitably artificial.
Ideally, we should judge new methods using real-world applications.
This is not feasible, but we use a realistic example by
downloading from the University of British Columbia library data centre
the {\it Survey of Labour and Income Dynamics} (SLID)
data provided by Statistics Canada (2014).
According to the read-me file, this survey complements traditional
survey data on labour market activity and income with an additional
dimension: the changes experienced by individuals over time.

We are grateful to Statistics Canada {for making} the data set available,
but we do not address the original goal of the survey here.
Instead, we use it as a superpopulation to study the effectiveness of our
small area quantile estimator.

After some data preprocessing, including removing units
containing missing values,
we retain 35488 sampling units and 6 variables. The variables are
\ttin, \gender, \age, \yrx, \tweek, and \edu,
i.e., total income, gender, age, years of experience,
number of weeks employed, and education level.
We transform $\ttin$ into $y = \log (2950+\ttin)$ so that its distribution
is closer to symmetric, where $2950$ is the $5$th percentile of
$\ttin$.
%%% New plots are needed, or simply do not provide any as
%%    our simulation is no longer design based.
We ignore the sampling plan under which this data set was obtained.
Instead, we examine how well our small area quantile
predictors perform if we sample from this ``real'' population.
We create 10 age groups:
\begin{center}
\tabcolsep 3pt  \small
\begin{tabular}{ccccccccccc } \hline
 $[0,\; 20)$  &  $[20,\; 25)$ &  $[25,\; 30)$& $[30,\; 35)$& $[35,\; 40)$&
 $[40,\; 45)$&  $[45,\; 50)$&  $[50,\; 55)$ &  $[55,\; 60)$ &  $[60,\; \infty)$   \\
 \hline
\end{tabular}
 \end{center}

Each age group is then divided into male and female subpopulations.
This gives a finite population with 20 small domains
(the small areas) based on age--gender combinations.
The sizes of these small {domains} are as follows.
\begin{center}
\tabcolsep 6pt
\begin{tabular}{|c|cccccccccc| } \hline
Male  & 1231 & 1525 &1372 &1337& 1469 & 1536& 1866& 1890&1920& 3089      \\
Female  & 1200 & 1433 &1449 &1504& 1497& 1695& 2053& 2019 &1944& 3459  \\  \hline
\end{tabular}
\end{center}

We first obtain the fitted values of the responses and residuals for
all the units under the standard NER model.
In each simulation repetition,  we create a shadow population
which keeps covariate $\bx_{kj}$ unaltered but assembles
new response value
\[
y_{kj} = \hat y_{kj} + \hat{\epsilon}_{k ,\pi(j)},
\]
where $\pi(\cdot)$ is random permutation of $\{1, \ldots, n_k\}$.
From this population,  we sample $n_k = 30$ units from area $k$ and
estimate the 5\%, 25\%, 50\%, 75\%, and 95\% small area quantiles
using NER, EL, MR, and EBEL.
For MR and EBEL, we assume that
the values of $\bx_{kj}$ are available for all units in the population.
We omit the other methods because our simulation studies showed that
they are less effective.

\begin{figure}
\begin{center}
\includegraphics[width=0.45\textwidth]{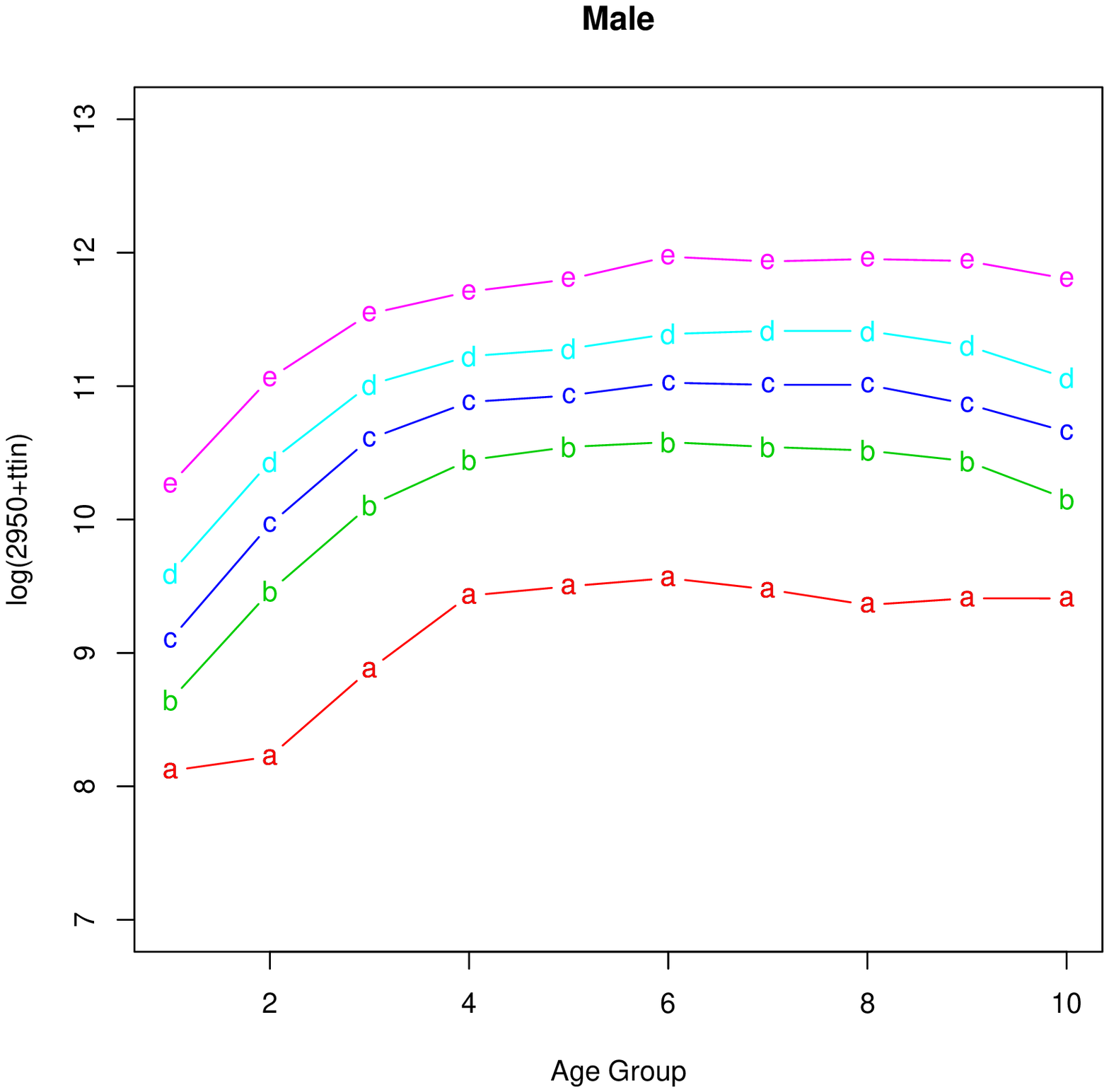}
\includegraphics[width=0.45\textwidth]{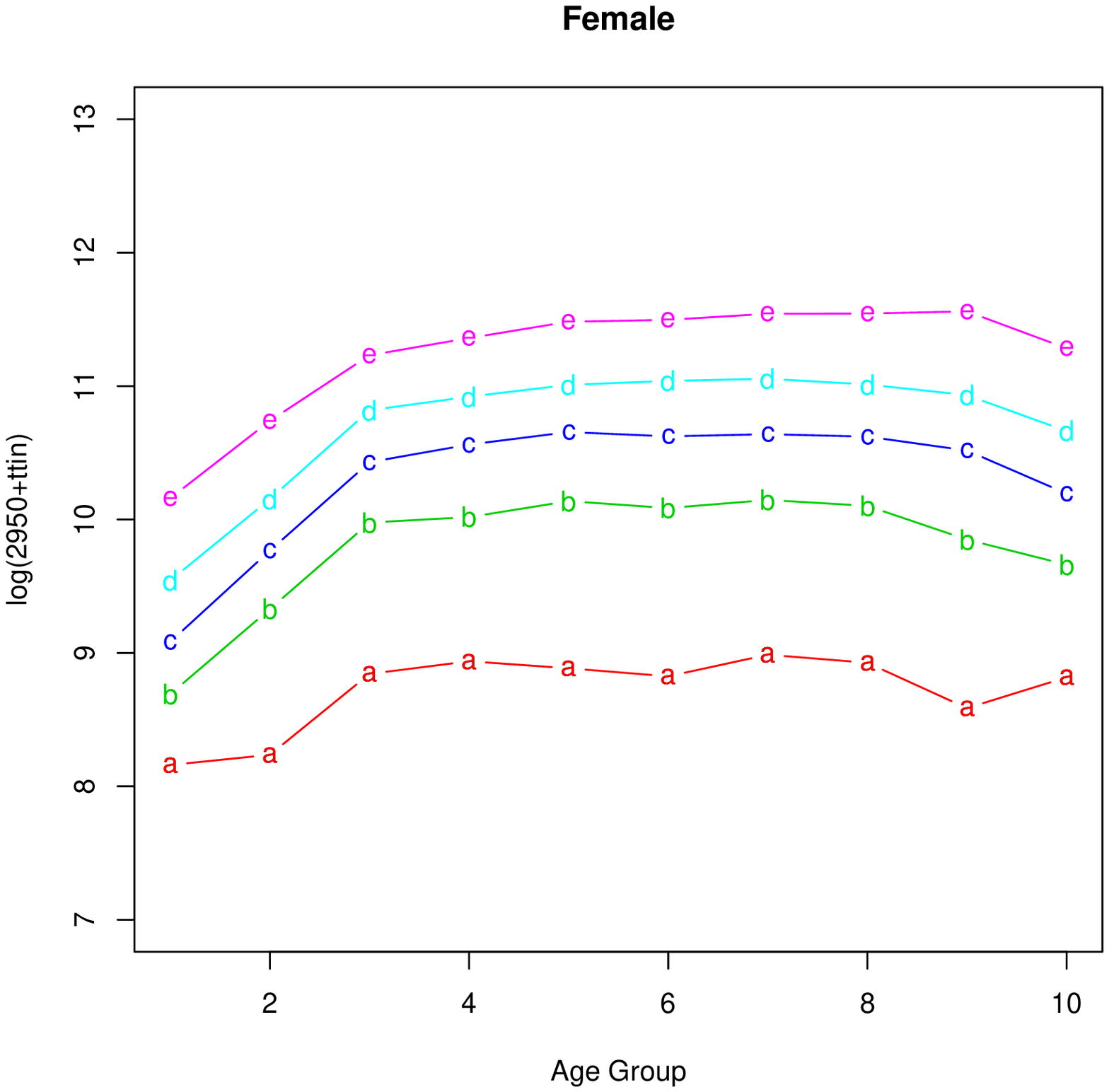}
\end{center}
\caption{Small area population quantiles for SLID data.
Lines a-d stand for area-specific 5\%, 25\%, 50\%, 75\% and 95\% quantiles, respectively.
\label{quan-realdata}}
\end{figure}

The population quantiles across the 10 age groups for both males and females
are displayed in  Figure \ref{quan-realdata}.
As expected,  total income increases as age increases for all quantiles
and both males and females.
We see that compared with the 95\% quantiles,
the 5\% quantiles for both males and females are much farther  from  the median.
Hence, the small area population distributions of the
response variable in all the small areas are skewed to the left.
It is harder to obtain accurate estimates for the lower quantiles than
for the upper quantiles.

We set the number of simulation repetitions to $500$.
The simulated \amse\ values and the ratio averages of the bootstrap
and simulated MSEs are given in Table \ref{table-realdata-nk=30}.
The proposed EL and EBEL quantile estimators clearly have
the best accuracy in terms of \amse.
Again, EL has surprisingly good performance,
although it does not use census information.
The performance of the bootstrap MSE estimates
for EL  and EBEL is satisfactory except for the 5\% quantiles.
This is likely due to the left skewness of the small area population
distribution.
The bootstrap MSE estimates work better for DIR than for MR.

\section{Conclusions and discussions}
We have proposed two general small area quantile estimation methods
under a nested error linear model: the NER under a normal
assumption on the error distribution and the EL under a DRM
assumption on the error distribution. They are applicable
whether or not census information on auxiliary variables is available.
Simulation shows that when the error distribution is not normal,
the DRM-based EL quantiles have superior performance.
The proposed resampling \amse\ estimates work reasonably well
for quantiles in the middle range.

\section*{Supplementary material}
The supplementary material contains proofs of Theorems 1--3
and some additional simulation results.

\renewcommand{\baselinestretch}{1}
\section*{References}
\begin{description}
\item
Anderson, J. A. (1979).
Multivariate logistic compounds.
{\it Biometrika} 66: 17--26.

\item
Ballini, F., Betti, G., Carrette, S. and Neri, L. (2006).
Poverty and inequality mapping in the Commonwealth of Dominica.
{\it Estudios Economicos} 2: 123--162.

\item
Battese, G. E., Harter, R. M. and Fuller, W. A. (1988).
An error-components model for prediction of county
crop areas using survey and satellite data.
{\it Journal of the American Statistical Association}
80: 28--36.

\item
Breckling, J. and Chambers, R. (1988).
M-quantiles.
{\it Biometrika} 75: 761--771.

\item
Chambers, R. and Dunstan, R. (1986).
Estimating distribution functions from survey data.
{\it Biometrika} 73: 597--604.

\item
Chambers, R. and Tzavidis, N. (2006).
M--quantile models for small area estimation.
{\it Biometrika} 93: 255--268.

\item
Chaudhuri, S. and Ghosh, M. (2011).
Empirical likelihood for small area estimation.
{\it Biometrika} 98: 473--480.

\item
Chambers, R., Chandra, H, Salvati, N and Tzavidis, N. (2011).
Outlier robust small area estimation.
{\it Journal of Royal Statistical Society, B} 76: 47--69.

%\item
%Chambers, R.,  Dreassi, E., and Salvati, N. (2014).
%Disease mapping via negative binomial regression M-q
%{\it Statistics in Medicine}  33(27): 4805--4824.

\item
Chen, J. and Liu, Y. (2013).
Quantile and quantile-function estimations under density ratio model.
{\it The Annals of Statistics}  4: 1669--1692.

%\item
%Chen, J., Rao, J.N.K. and Sitter, R.R. (2000).
%Efficient random imputation for missing data in complex surveys.
%{\it Statistica Sinica}  10: 1153--1169.

%\item
%Chen, J. and Wu, C. (2002).
%Estimation of distribution function and quantiles using
%the model-calibrated pseudo empirical likelihood method.
%{\it Statistica Sinica} 12: 1223--1239.

\item
Diallo, M. S. and Rao, J.~N.~K. (2016).
Small area estimation of complex parameters under unit-level
models with skew-normal errors.
Manuscript.

\item
Elbers, C. Lanjouw, J. O. and Lanjouw, P. (2003).
Micro-level estimation of poverty and inequality.
{\it Econometrica} 71: 355--364.

\item
Estevao, V. M. and S\'{a}rndal, C. E. (2006).
Survey estimates by calibration on complex auxiliary information.
{\it International Statistical Review} 74: 127--147.

\item
Fay, R. E. and Herriot, R. A. (1979).
Estimates of income for small places:
An application of James-Stein procedures to census data.
{\it Journal of the American Statistical Association}
 74: 269--277.

%\item
%Fokianos, K., Kedem, B., Qin, J. and Short, D. A. (2001).
%A semiparametric approach to the one-way layout.
%{\it Technometrics}, {\bf 43}, 56-65.

%\item
%Francisco, C. A. and Fuller, W. A. (1991).
%Quantiles estimation with a complex survey design.
%{\it The Annals of Statistics}  19: 454--469

\item
Ghosh, M., Maiti, T. and Roy, A. (2008).
Influence functions and robust Bayes and empirical Bayes small area estimation.
{\it Biometrika}  95: 573--585.

\item
Guadarrama, M., Molina, I. and Rao, J.~N.~K. (2016).
Small area estimation of general parameters under complex sampling designs.
Manuscript.

\item
Haslett, S. and Jones, G. (2005).
Small area estimation using surveys and some practical and statistical issues.
{\it Statistics in Transition} 7: 541--555.

\item
Jiang, J. and Lahiri, P. S. (2006).
Estimation of finite population domain means: A
model-assisted empirical best prediction approach.
{\it Journal of the American Statistical Association}
101: 301--311.

\item
Jiang, J.  and Nguyen, T. (2012).
Small area estimation via heteroscedastic nested-error regression.
{\it The Canadian Journal of Statistics}
40: 588--603.

\item
Jiang, J.,  Nguyen, T. and Rao, J. S. (2010).
Fence method for nonparametric small area estimation.
{\it Survey Methodology}  36: 3--11.

\item
Jiongo, V. D., Haziza, D. and Duchesne, P. (2013).
Controlling the bias of robust small-area estimators.
{\it Biometrika} 100: 843--858.

\item
Keziou, A. and  Leoni-Aubin, S. (2008).
On empirical likelihood for semiparametric two-sample density
ratio models.
{\it Journal of Statistical Planning and Inference}  138: 915--928.

\item
Koenker R. and Bassett, G. (1978).
Regression quantiles. {\it Econometrica} 46: 33--50.

\item
Kriegler, B.  and Berk, R. (2010).
Small area estimation of the homeless in Los Angeles:
An application of cost-sensitive stochastic
gradient boosting.
{\it The Annals of Applied Statistics}
4: 1234--1255.

\item
Lahiri, P. S. and Rao, J.~N.~K. (1995).
Robust estimation of mean squared error of small
area estimators.
{\it Journal of the American Statistical Association}
 90: 758--766.

\item
Marchetti, S., Tzavidis, N. and Pratesi, M. (2012).
Non-parametric bootstrap mean squared error estimation for
M-quantile estimators of small area averages, quantiles and
poverty indicators.
{\it Computational Statistics and Data Analysis} 56:
2889--2902.

\item
Molina, I. and Rao, J. N. K. (2010).
Small area estimation of poverty indicators.
{\it The Canadian Journal of Statistics} 38: 369--385.

\item
Neri, L., Ballini F. and Betti, G. (2005).
Poverty and inequality in transition countries.
{\it Statistics in Transition} 7: 135--157.

\item
Opsomer, J.~D., Claeskens, G., Ranalli, M.~G., Kauermann, G. and Breidt, F.~J. (2008).
Non-parametric small area estimation
using penalized spline regression.
{\it Journal of the Royal Statistical Society: B}
70: 265--286.

\item
Owen, A. B. (1988).
Empirical likelihood ratio confidence
intervals for a single functional.
{\it Biometrika}  75: 237--249.

\item
Owen, A. B. (2001).
{\it Empirical Likelihood}. New York: Chapman and Hall/CRC.

\item
Pfeffermann, D. (2002).
Small area estimation: New developments and directions.
{\it International Statistical Review} 70: 125--143.

\item
Pfeffermann, D. (2013).
New important developments in small area estimation.
{\it Statistical Science} 28: 40--68.

\item
Pfeffermann, D. and Sverchkov, M. (2007).
Small-area estimation under informative
probability sampling of areas and within the selected areas.
{\it Journal of the American Statistical Association}
102: 1427--1439.

\item
Prasad, N. G. N. and Rao, J.~N.~K. (1990).
The estimation of mean squared errors of small area estimators.
{\it Journal of the American Statistical Association}
85: 163--171.

\item
Qin, J. (1998).
Inferences for case-control and semiparametric two-sample
density ratio models.
{\it Biometrika}  85: 619--630.

\item
Qin, J. and Lawless, J. F. (1994).
 Empirical likelihood and general estimating equations.
{\it Annals of Statistics} 22: 300--325.

\item
Qin, J. and Zhang, B. (1997).
A goodness-of-fit test for logistic regression models based on
case-control data.
{\it Biometrika} 84: 609--618.

\item
Ranalli, M. G., Breidt, F. J. and Opsomer, J. D. (2016).
Nonparametric regression methods for small area estimation.
In Pratesi, M. (ed.),
{\it Analysis of Poverty Data by Small Area Estimation}, New York: Wiley, pp. 187--204.

\item
Rao, J. N. K. (2003). {\it Small Area Estimation}.  New York: Wiley.

\item
Rao, J. N. K. and Molina, I. (2015).
{\it Small Area Estimation}. Hoboken, NY: Wiley.

%\item
%Rao, J. N. K., Kovar, J. G. and Mantel, H. J. (1990).
%On estimating distribution functions
%and quantiles from survey data using auxiliary information.
%{\it Biometrika}  {\bf 77}, 365-375.

%\item
%Rao, J. N. K., Hartley, H. O. and Cochran, W. G. (1962).
%On a simple procedure of unequal probability sampling without replacement.
%{\it Journal of the Royal Statistical Society}, Ser. B, {\bf 24}, 482-491.

%\item
%R Development Core Team (2011).
%{\it R: A language and environment for statistical computing}.
%R Foundation for Statistical Computing, Vienna, Austria.

%\item
%Rao, J. N. K. and Wu, C.-F. J. (1988).
%Resampling inference with complex survey data.
%{\it Journal of the American Statistical Association},
%{\bf 83}, 231-241.

\item
Schaible, W. L. (1993).
Use of small area estimators in U.S. federal programs.
In Kalton, G., Kordos, J. and Platek, R. (eds), {\it Small Area Statistics and Survey Designs},
Warsaw: Central Statistical Office, I: 95--114.

\item
Statistics Canada (2014). {\it Survey of labour and income dynamics, 2011}.
Access: ABACUS. \url{http://hdl.handle.net/10573/42961}.

\item
Tarozzi, A. and Deaton, A. (2009).
Using census and survey data to estimate poverty and inequality for small
areas.
{\it Review of Economics and Statistics} 91: 773--792.

\item
Tzavidis, N. and Chambers, R. (2005).
Bias adjusted estimation for small areas with
M-quantile models.
{\it Statistics in Transition} 7: 707--713.

%\item
%Tzavidis, N., Marchetti, S. and Chambers, R. (2010).
%Robust prediction of small area means and distributions.
%Aust. \& NZ J. Stat., 52, 167-186.

\item
Tzavidis, N., Salvati, N., Pratesi, M. and Chambers, R. (2008).
M-quantile models with application to poverty mapping.
{\it Statistical Methodology and Applications} 17: 393--411.

%\item
%van der Vaart, A. W. and  J. A. Wellner (1996).
%{\it Weak Convergence and Empirical Processes}.
%New York: Springer.

\item
Verret, F., Rao, J.~N.~K. and Hiridoglou, M.~A. (2015).
Model-based small area estimation under informative sampling.
{\it Survey Methodology} 41: 333--347.

%\item
%Wang, S. and Dorfman, A.H. (1996).
%A new estimator for the finite population distribution
%function.
%{\it Biometrika}  83, 639--652.

\item
You, Y. and Rao, J.~N.~K. (2002).
A pseudo-empirical best linear unbiased predictor
approach to small area estimation using survey weights.
{\it Canadian Journal of Statistics} 30: 431--439.

\item
Zhang, B. (1997).
Assessing goodness-of-fit of generalized logit models
based on case-control data.
{\it Journal of Multivariate Analysis} 82: 17--38.

\item
Zhang, B. (2000).
Quantile estimation under a two-sample semi-parametric model.
{\it Bernoulli} 6: 491--511.
\end{description}

\begin{table}[ht]
\caption{\amse\  of small area quantile estimators under model \eqref{model1}}
\centerline{{\small Sample size $n=600$, number of repetitions 1000, $\bbeta = 1.5\bbeta_0$}}
\label{table-amse-nk=30}
 \vspace{1ex}
\begin{center}
\tabcolsep 12pt
\renewcommand{\arraystretch}{1.2} \small
 \vspace{-0.4cm}
\begin{tabular}{cl|rrrrr }
 \hline
  &  &\multicolumn{5}{|c }{AMSE}       \\

\hline
Scenario & $\alpha$ &  $5\%$ & $25\%$   & $50\%$  &  $75\%$    & $95\%$  \\  \hline
 (i)    & DIR      &  0.4242 &0.1490 &0.1244 &0.1499 &0.4324  \\
        & NER      &  0.0806 &0.0659 &0.0633 &0.0656 &0.0802  \\
        & EL       &  0.0878 &0.0709 &0.0682 &0.0705 &0.0875  \\ [1ex]
        & MQ       &  0.1926 &0.0920 &0.0764 &0.0929 &0.2021  \\
        & MR       &  0.0774 &0.0657 &0.0634 &0.0650 &0.0765  \\
        & EB       &  0.0797 &0.0680 &0.0660 &0.0676 &0.0789  \\
        & EBEL     &  0.0861 &0.0729 &0.0709 &0.0724 &0.0852  \\ \hline

 (ii)   & DIR      &  0.3234 &0.1404 &0.1236 &0.1405 &0.3130   \\
        & NER      &  0.0753 &0.0620 &0.0569 &0.0615 &0.0741   \\
        & EL       &  0.0841 &0.0695 &0.0667 &0.0690 &0.0829   \\ [1ex]
        & MQ       &  0.1376 &0.0819 &0.0747 &0.0823 &0.1402   \\
        & MR       &  0.0708 &0.0603 &0.0571 &0.0600 &0.0704   \\
        & EB       &  0.0729 &0.0629 &0.0590 &0.0628 &0.0722   \\
        & EBEL     &  0.0805 &0.0711 &0.0691 &0.0709 &0.0799   \\ \hline

 (iii)  & DIR      &  0.7323 &0.1634 &0.0977 &0.1025 &0.2597   \\
        & NER      &  0.2034 &0.0821 &0.0712 &0.0576 &0.1118   \\
        & EL       &  0.1303 &0.0573 &0.0521 &0.0540 &0.0681   \\ [1ex]
        & MQ       &  0.4028 &0.1162 &0.0567 &0.0641 &0.1607   \\
        & MR       &  0.1756 &0.0852 &0.0699 &0.0560 &0.1206   \\
        & EB       &  0.1950 &0.0848 &0.0737 &0.0594 &0.1146   \\
        & EBEL     &  0.1284 &0.0572 &0.0539 &0.0549 &0.0633   \\ \hline

 (iv)   & DIR      &  0.2621 &0.1028 &0.0975 &0.1627 &0.7385  \\
        & NER      &  0.1138 &0.0589 &0.0720 &0.0835 &0.2060  \\
        & EL       &  0.0684 &0.0551 &0.0529 &0.0584 &0.1313  \\ [1ex]
        & MQ       &  0.0983 &0.0518 &0.0534 &0.1020 &0.4117  \\
        & MR       &  0.1228 &0.0572 &0.0708 &0.0870 &0.1774  \\
        & EB       &  0.1169 &0.0606 &0.0746 &0.0866 &0.1970  \\
        & EBEL     &  0.0636 &0.0560 &0.0549 &0.0586 &0.1291  \\
 \hline
\end{tabular}
\end{center}
\end{table}

\begin{table}[ht]
\caption{  Average ratios of estimated and simulated MSEs under model \eqref{model1}}
\centerline{{\small Sample size $n=600$,    $B=100$,
$\bbeta = 1.5\bbeta_0$,  number of repetitions 1000}}
\label{table-ratio-nk=30}
\begin{center}
\tabcolsep 10pt
\renewcommand{\arraystretch}{1.2} %\small
% \vspace{-0.6cm}
\begin{tabular}{clccccc }
 \hline
Scenario& $\alpha$ &  $5\%$ & $25\%$   & $50\%$  &  $75\%$    & $95\%$      \\  \hline
 (i)    & DIR      &  0.9693& 0.9835& 0.9892& 0.9780& 0.9535  \\
        & EL       &  0.9307& 0.9536& 0.9607& 0.9546& 0.9322  \\[1ex]
        & MR       &  0.9784& 0.9823& 0.9901& 0.9904& 0.9906  \\
        & EBEL     &  0.9574& 0.9686& 0.9732& 0.9716& 0.9686  \\ \hline

 (ii)   & DIR      &  0.9819& 0.9663& 0.9683& 0.9774& 0.9939   \\
        & EL       &  0.8830& 0.9150& 0.9411& 0.9296& 0.9016   \\[1ex]
        & MR       &  0.9525& 0.9503& 0.9769& 0.9586& 0.9637   \\
        & EBEL     &  0.9143& 0.9310& 0.9503& 0.9381& 0.9243   \\ \hline

 (iii)  & DIR      &  0.9564& 0.9463& 0.9915& 0.9977& 0.9845   \\
        & EL       &  0.7006& 0.9769& 0.9787& 0.9739& 0.9585   \\[1ex]
        & MR       &  0.3874& 0.6753& 0.8014& 1.0265& 0.5598   \\
        & EBEL     &  0.7430& 0.9830& 0.9817& 0.9800& 0.9757   \\ \hline

 (iv)   & DIR      &  0.9723& 0.9938& 0.9918& 0.9505& 0.9541   \\
        & EL       &  0.9549& 0.9466& 0.9523& 0.9568& 0.6942   \\[1ex]
        & MR       &  0.5508& 1.0016& 0.7882& 0.6631& 0.3841   \\
        & EBEL     &  0.9733& 0.9563& 0.9564& 0.9538& 0.7399   \\
 \hline
\end{tabular}
\end{center}
 \end{table}

\begin{figure}[ht]
\caption{\small Area-specific MSEs  (upper half of each plot) and
ratios of bootstrap and simulated MSEs  (lower half   of each plot) for
Scenarios (i) and (iv).
In this setting, sample size  $n=600$,
number of bootstrap repetitions $B=100$, and  $\bbeta = 1.5\bbeta_0$.}
\label{fig-s1}   \vspace{-3ex}
\begin{center}
\begin{tabular}{cc}
\includegraphics[width=0.45\textwidth]{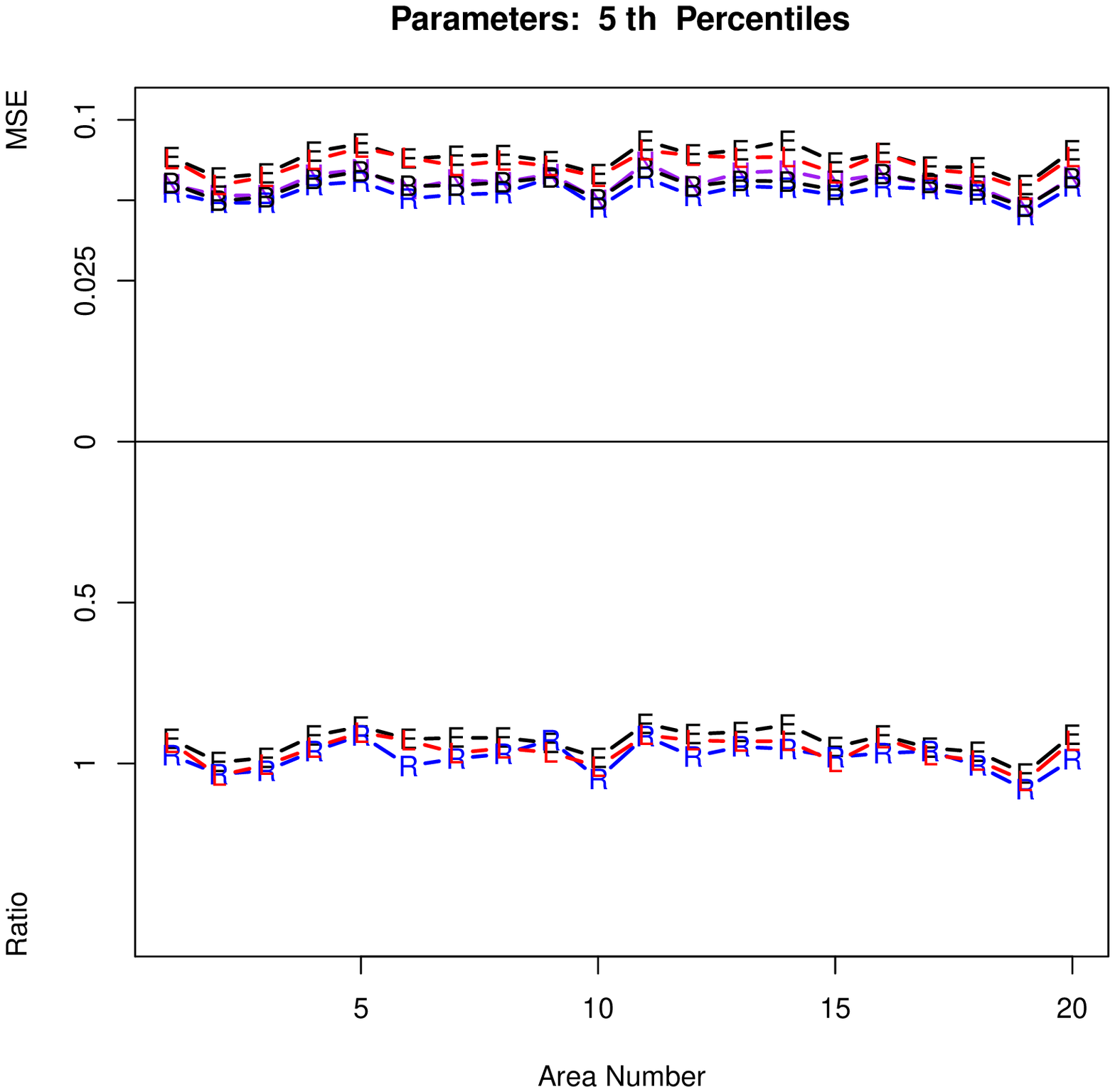}&
\includegraphics[width=0.45\textwidth]{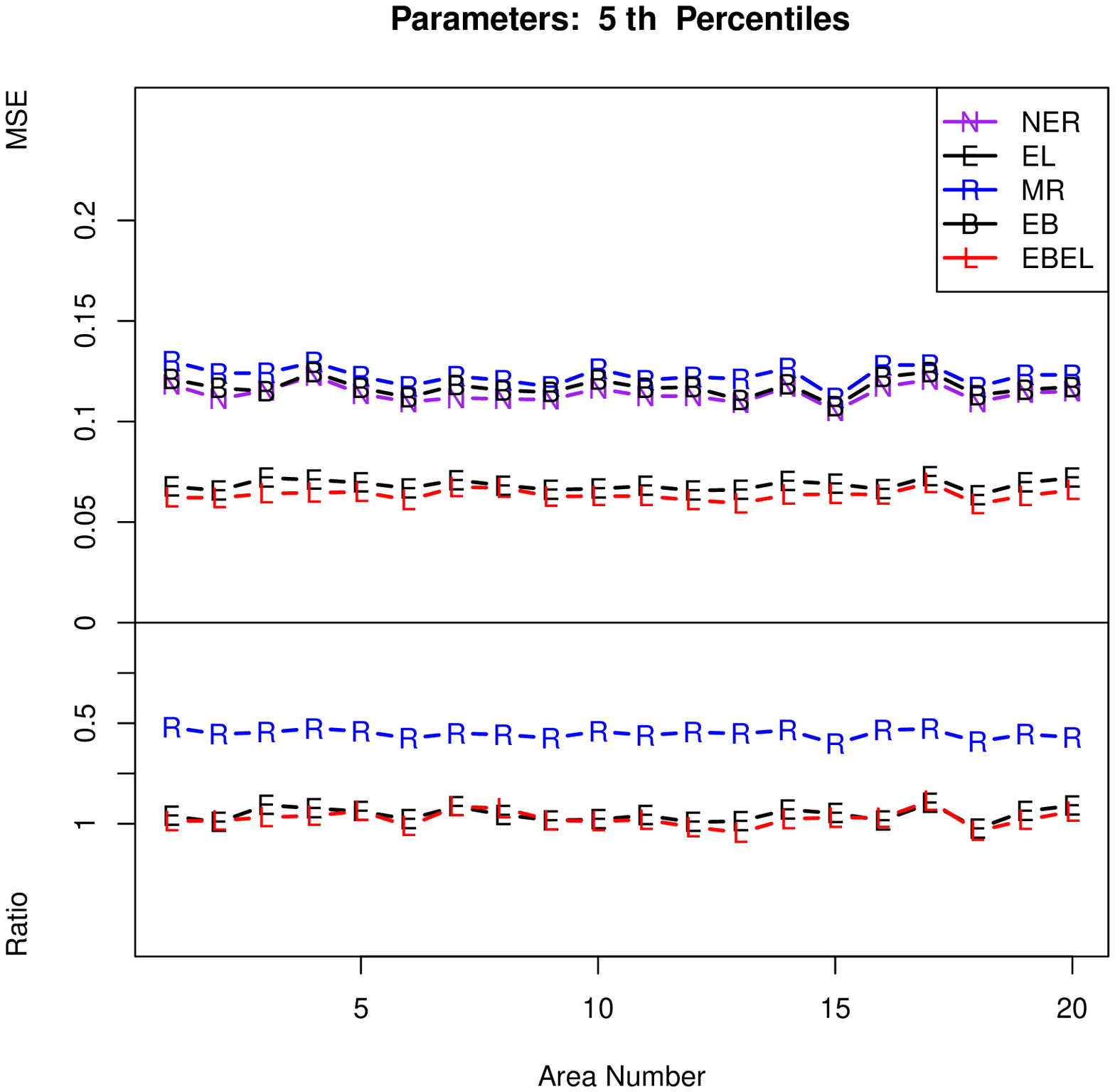}\\
\includegraphics[width=0.45\textwidth]{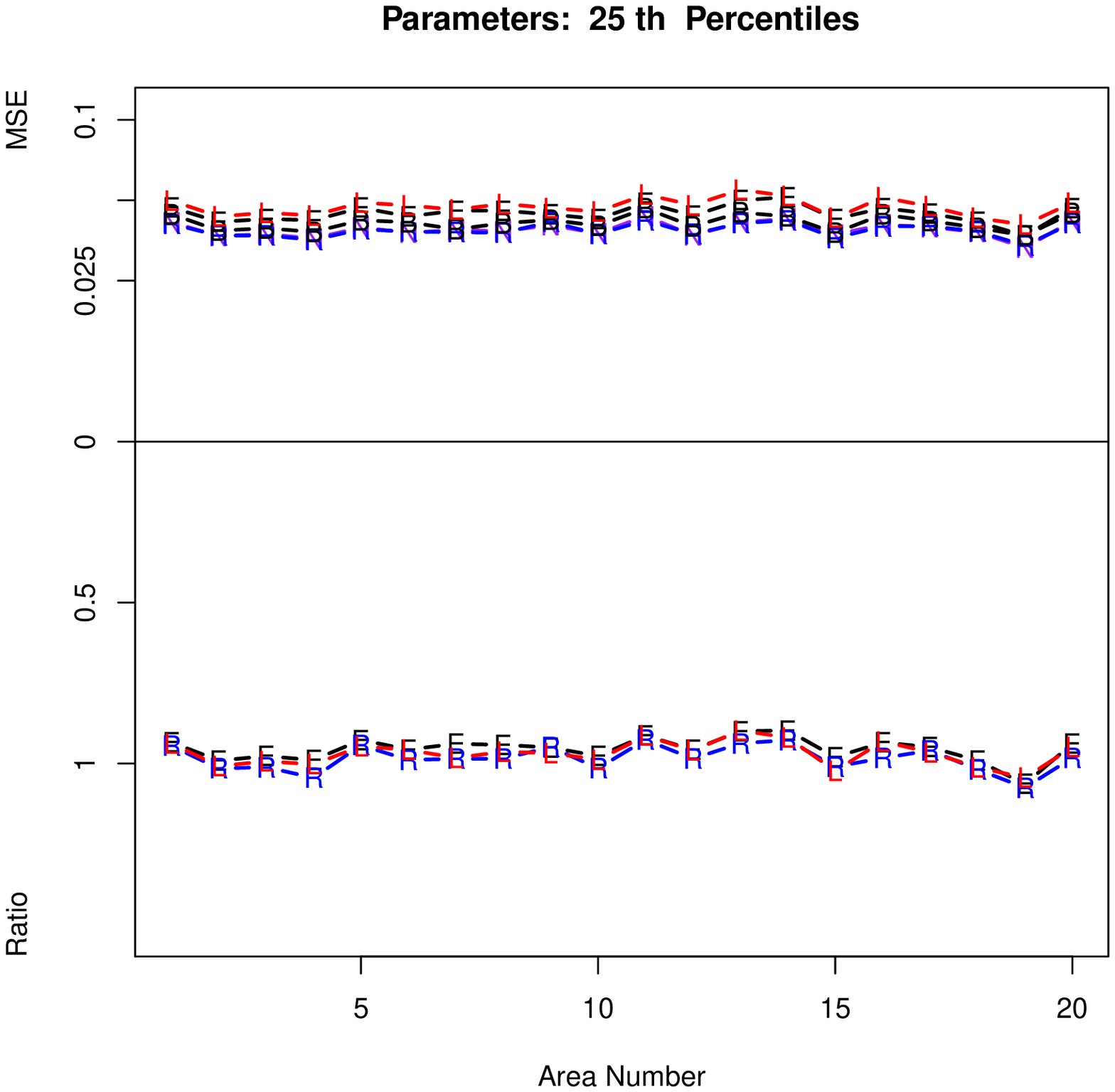}&
\includegraphics[width=0.45\textwidth]{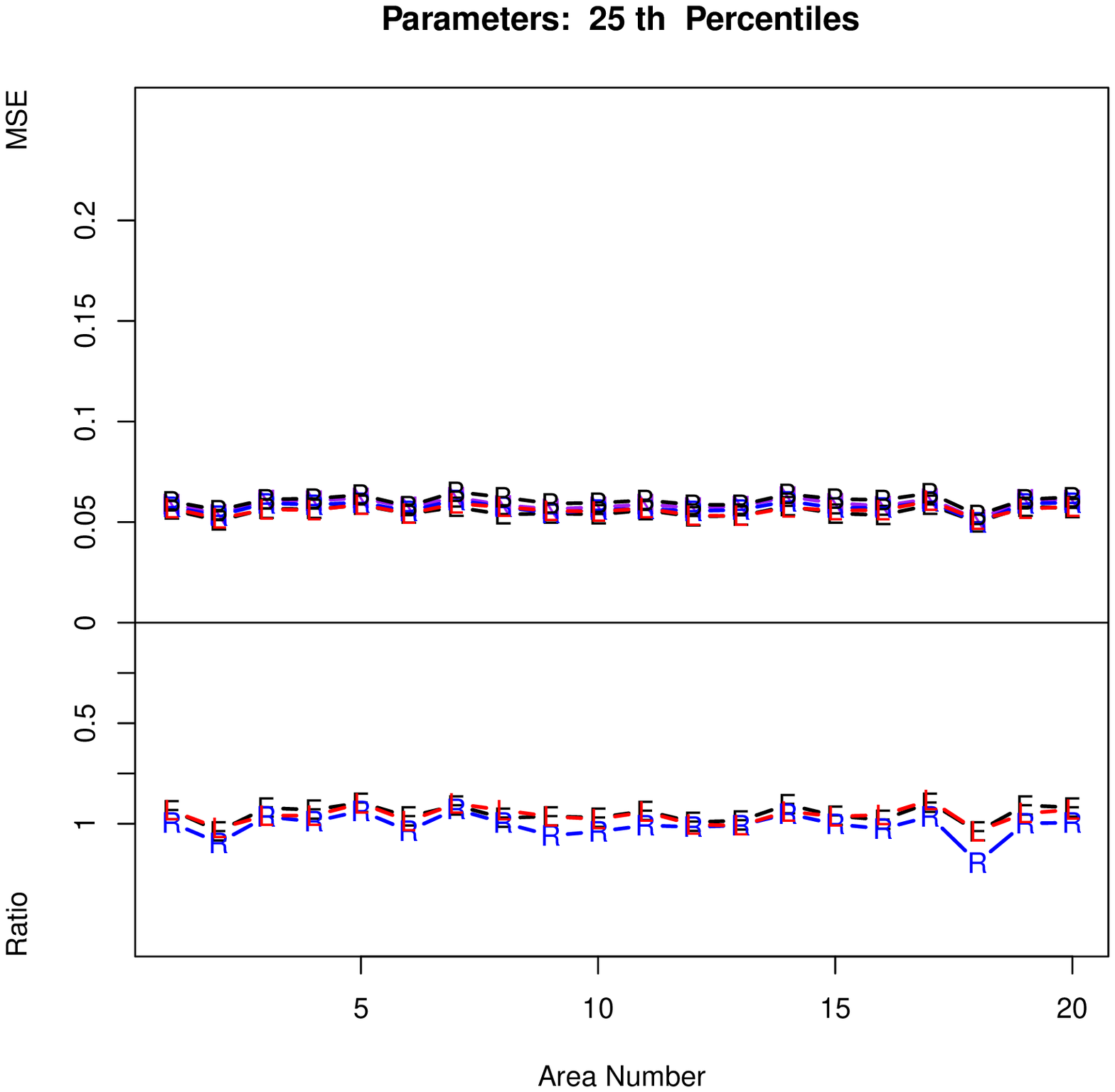}\\
\includegraphics[width=0.45\textwidth]{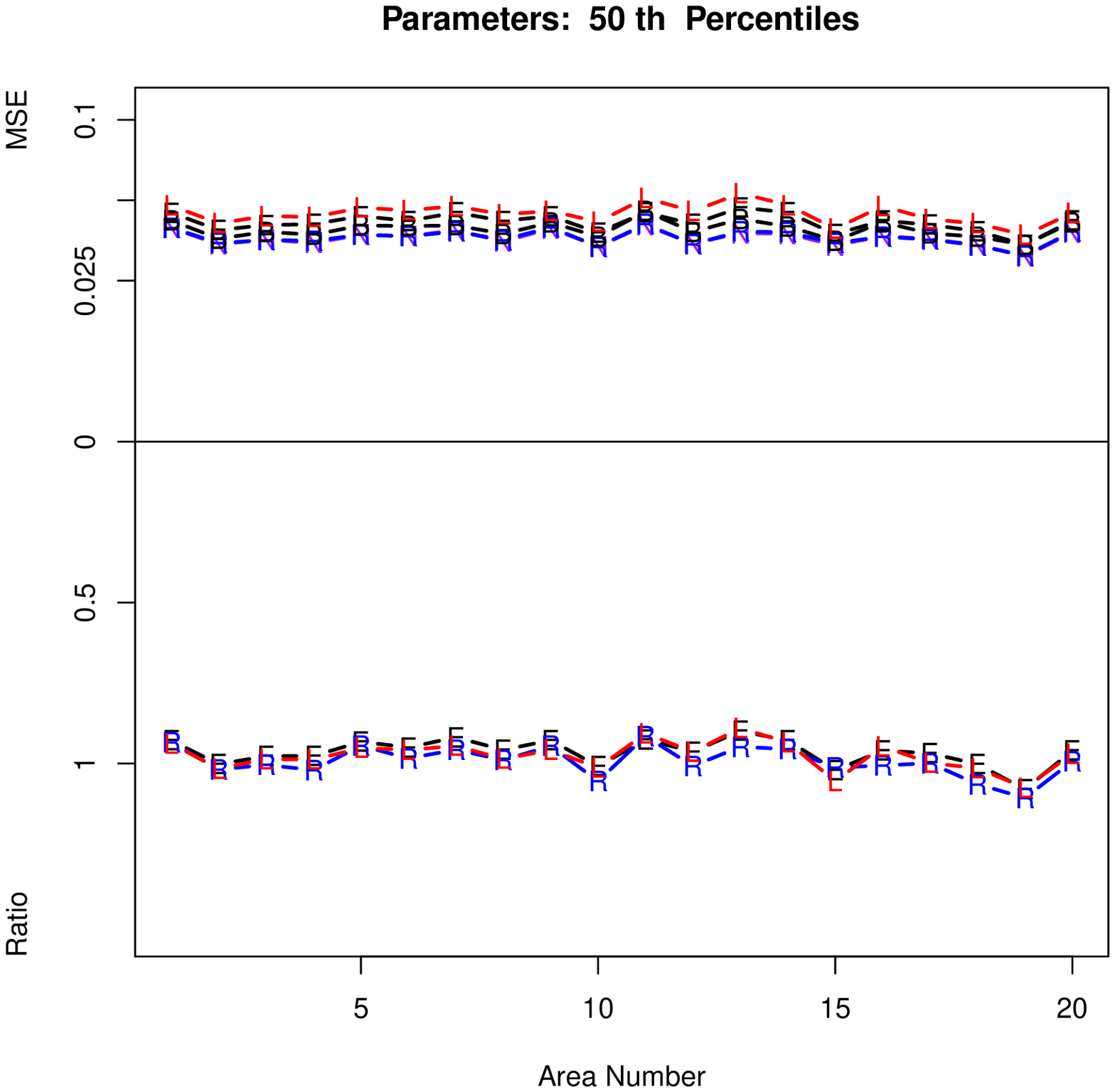}&
\includegraphics[width=0.45\textwidth]{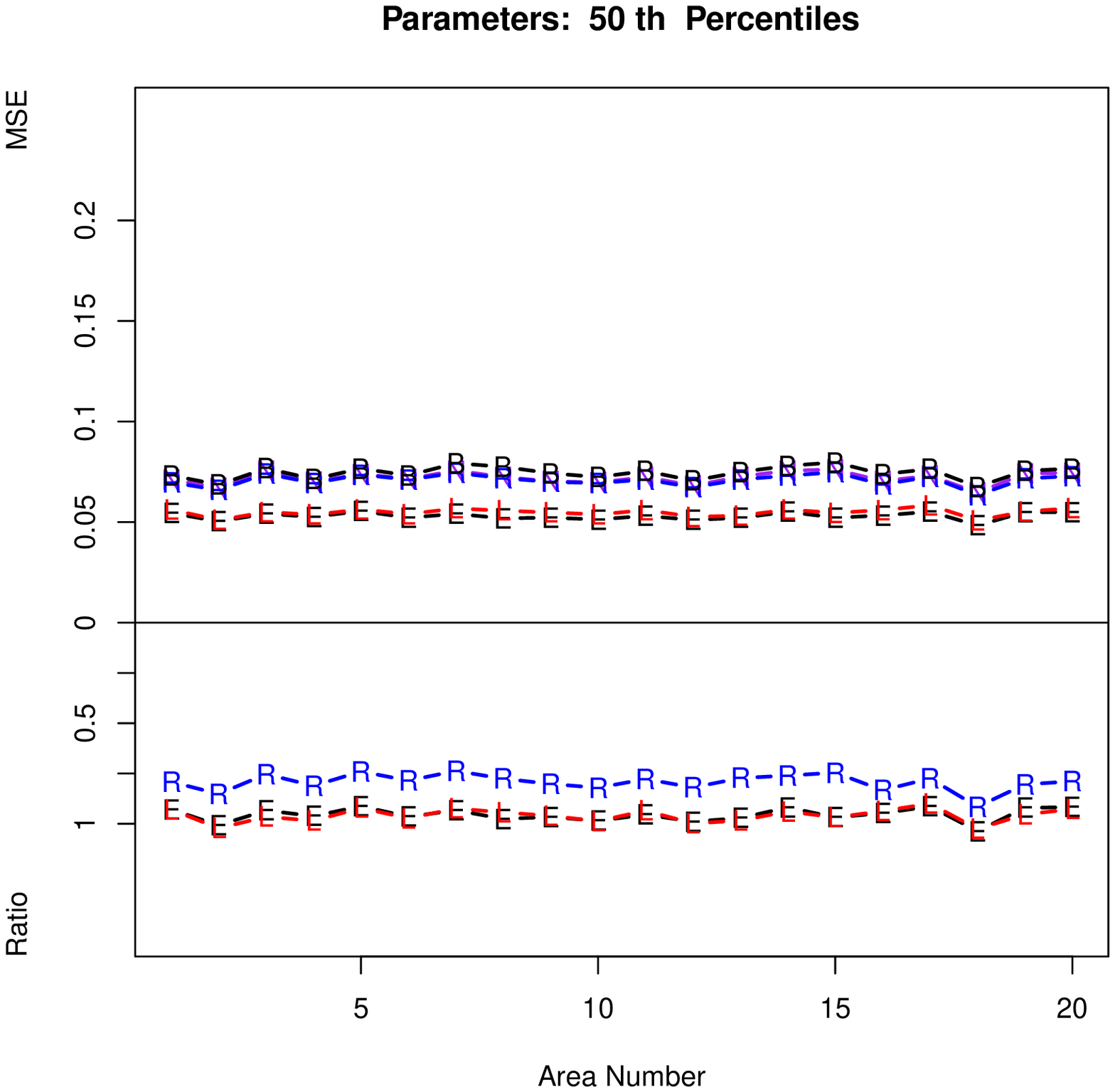}
\end{tabular}
\end{center}
\end{figure}

\begin{figure}[ht]
\caption{\small (continued) Area-specific MSEs  (upper half of each plot) and
ratios of bootstrap and simulated MSEs  (lower half   of each plot) for
Scenarios (i) and (iv).
In this setting, sample size  $n=600$,
number of bootstrap repetitions $B=100$, and  $\bbeta = 1.5\bbeta_0$.}
\label{fig-s2}
\begin{center}     \vspace{-3ex}
\begin{tabular}{cc}
\includegraphics[width=0.45\textwidth]{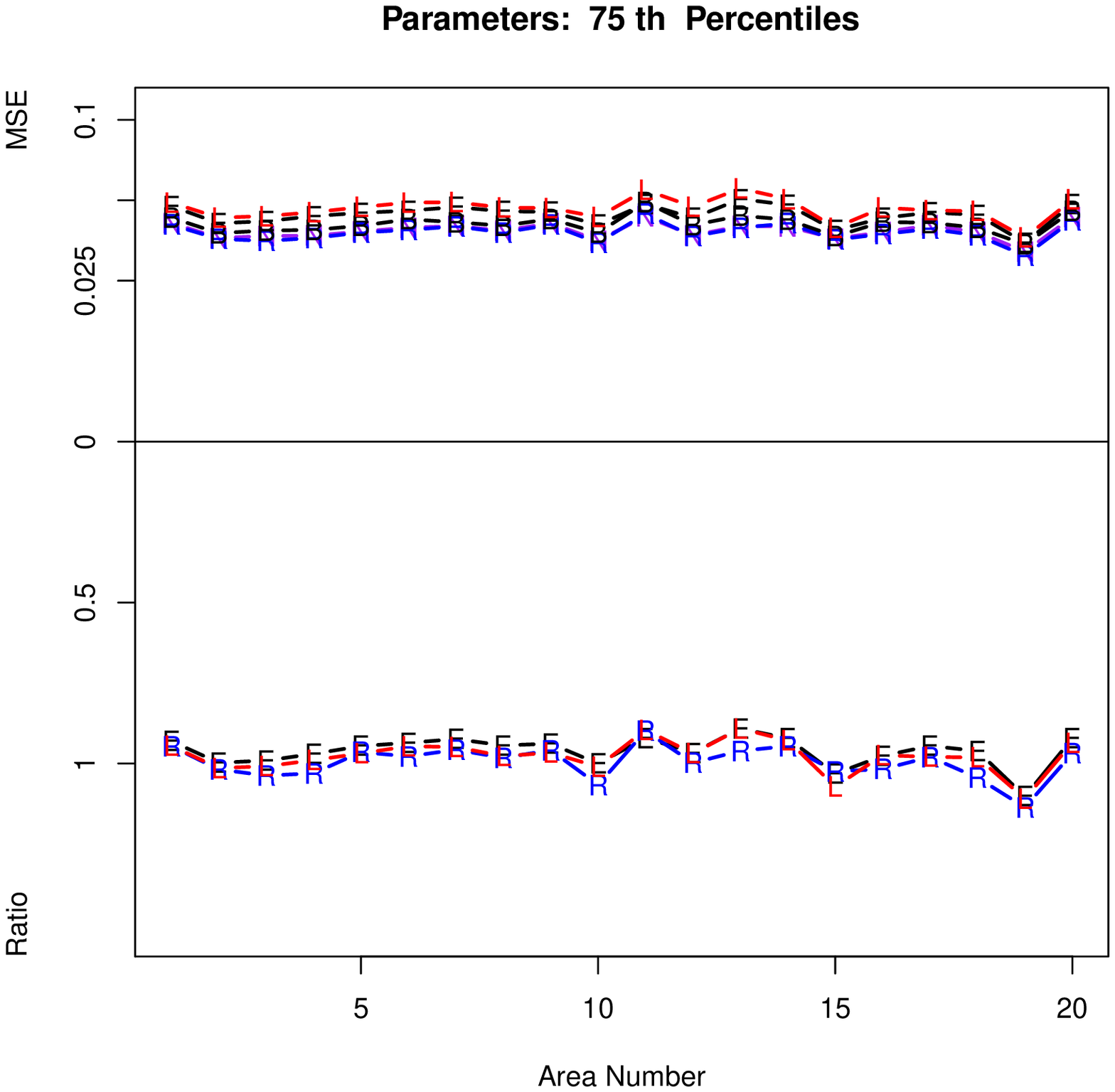}&
\includegraphics[width=0.45\textwidth]{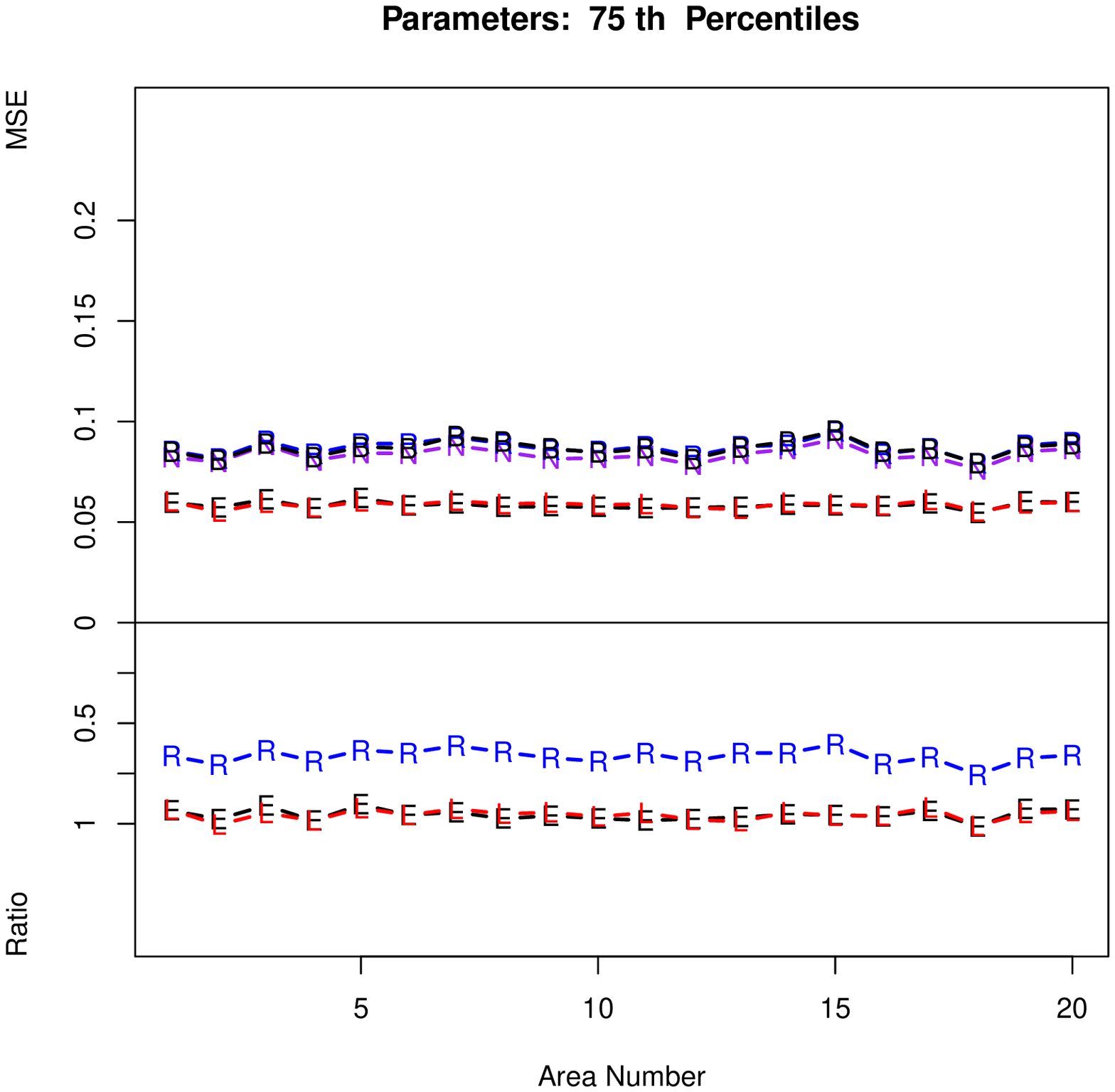}\\
\includegraphics[width=0.45\textwidth]{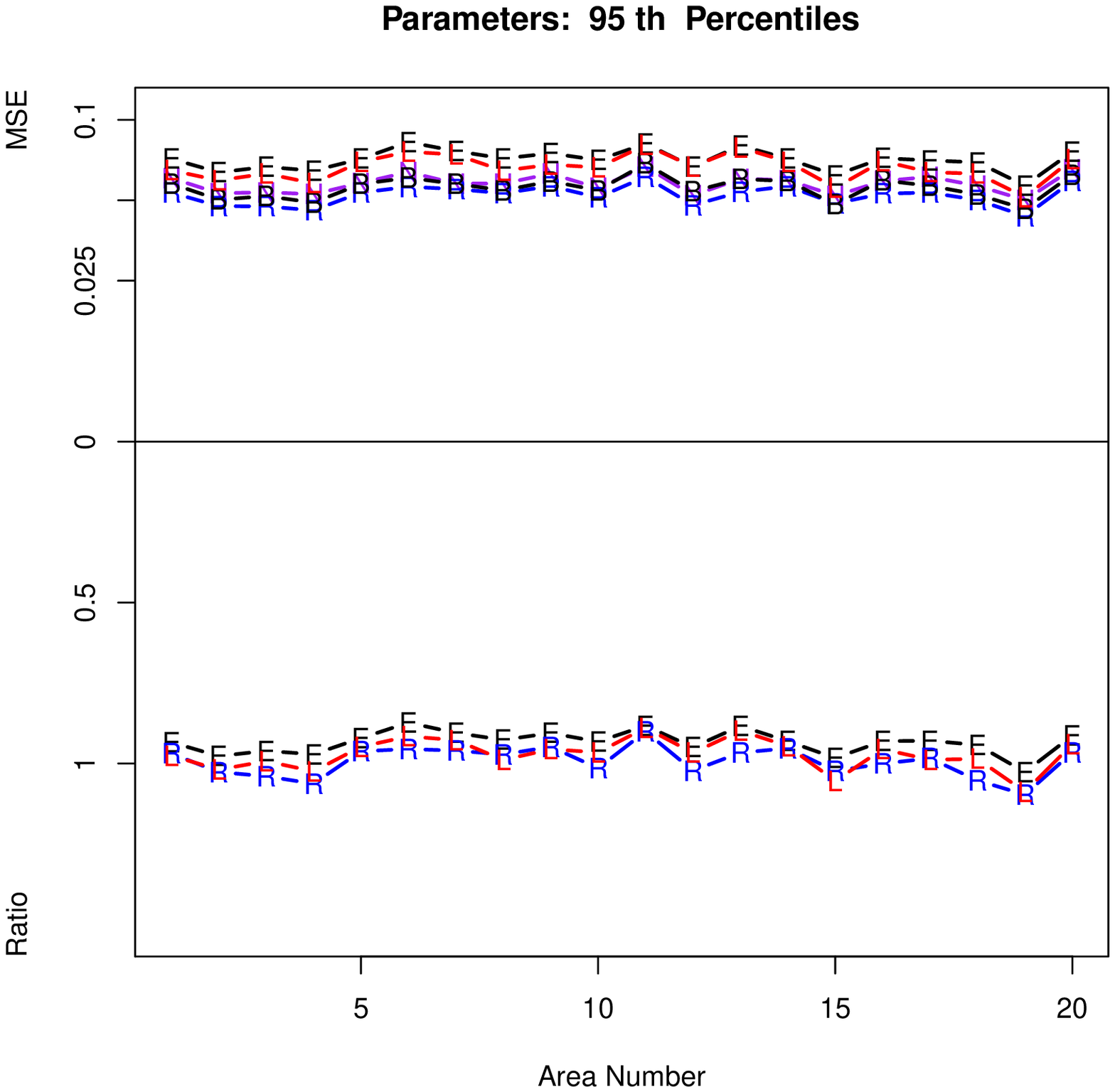}&
\includegraphics[width=0.45\textwidth]{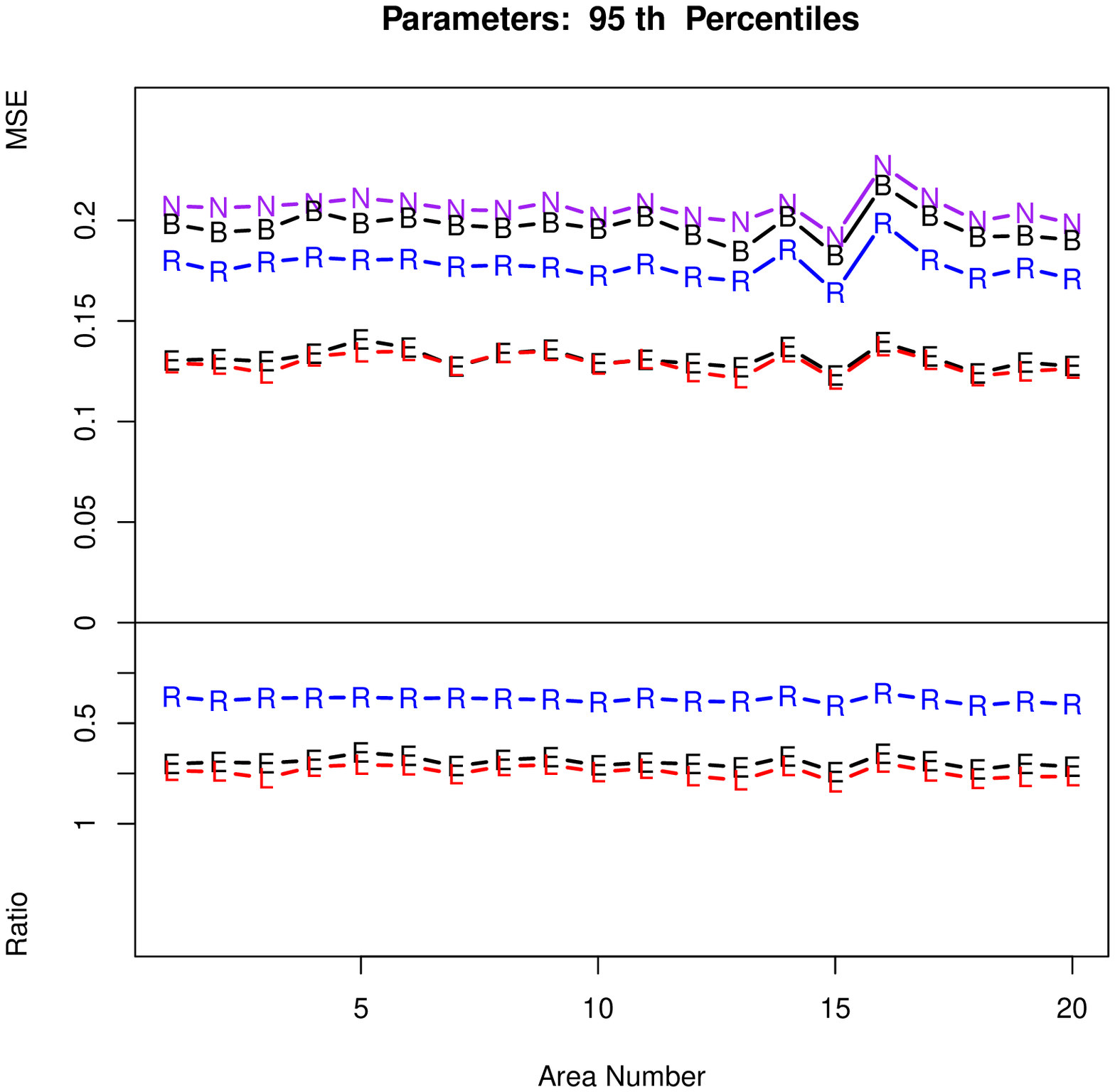}
\end{tabular}
\end{center}
\end{figure}

\begin{table}[ht]
\caption{ Simulation results of small area quantile estimators based on SLID data}
\centerline{Sample size $n=600$, $B=100$ for bootstrap, number of repetitions $500$.}
\label{table-realdata-nk=30}
\begin{center}
\tabcolsep 12pt
\renewcommand{\arraystretch}{1.2} \small
 \vspace{-0.4cm}
\begin{tabular}{l rrrrr }
 \hline
   &\multicolumn{5}{ c }{AMSE}       \\
\hline
 $\alpha$ &  $5\%$ & $25\%$   & $50\%$  &  $75\%$    & $95\%$  \\  \hline
 DIR      &  0.1903& 0.0455& 0.0208& 0.0201& 0.0882   \\
 NER      &  0.0709& 0.0259& 0.0205& 0.0165& 0.0419   \\
 EL       &  0.0712& 0.0153& 0.0136& 0.0141& 0.0205   \\ [1ex]
 MQ       &  0.1144& 0.0347& 0.0141& 0.0259& 0.1011   \\
 MR       &  0.0573& 0.0258& 0.0197& 0.0157& 0.0438   \\
 EB       &  0.0689& 0.0246& 0.0188& 0.0160& 0.0430   \\
 EBEL     &  0.0712& 0.0150& 0.0131& 0.0140& 0.0212   \\   \hline
    &\multicolumn{5}{ c }{Ratio of bootstrapped and simulated MSEs}       \\
\hline
 $\alpha$ &  $5\%$ & $25\%$   & $50\%$  &  $75\%$    & $95\%$  \\  \hline
 DIR      &  1.1722& 0.8961& 1.1356& 1.1552& 1.1412 \\
 EL       &  0.4501& 0.9066& 0.9202& 0.9966& 0.8805 \\
 MR       &  0.4063& 0.5612& 0.6694& 1.0420& 0.4576 \\
 EBEL     &  0.3462& 0.8469& 0.9143& 0.9544& 0.8115 \\   \hline
\end{tabular}
\end{center}
\end{table}

%%% Hopefully, the result on bootstrap variance estimation will be better.

\end{document}